\documentclass[aps,prd,amsmath,amssymb,reprint,superscriptaddress,nofootinbib]{revtex4-2}

\usepackage[colorlinks=true,allcolors=blue]{hyperref}
\usepackage{graphicx}
\usepackage{dcolumn}
\usepackage{bm}
\usepackage[mathlines]{lineno}
\usepackage[shortlabels]{enumitem}
\usepackage{lipsum}
\usepackage{xspace}
\usepackage[nolist]{acronym}
\usepackage{color}
\usepackage[caption=false]{subfig}
\usepackage{bm}
\usepackage[normalem]{ulem}
\usepackage{verbatim}
\usepackage{xcolor}
\usepackage{booktabs}

\graphicspath{{images/}}

\newcommand\ifnotSI[1]{}

\usepackage{etoolbox}
\makeatletter
\patchcmd\linenumberpar{\@LN@parpgbrk}{\penalty\@LN@parpgpen\relax}{}{}
\makeatother

\makeatletter
\newcommand{\customlabel}[2]{%
   \protected@write \@auxout {}{\string \newlabel {#1}{{#2}{\thepage}{#2}{#1}{}} }%
   \hypertarget{#1}{#2}
}
\makeatother

\begin{document}

\title{Electromagnetic modeling and science reach of DMRadio-m$^3$}

\author{A.~AlShirawi}
\affiliation{Department of Physics, Stanford University, Stanford, CA 94305}

\author{V.~Ankel}
\affiliation{Department of Physics, Stanford University, Stanford, CA 94305}
\affiliation{Kavli Institute for Particle Astrophysics and Cosmology, Stanford University, Stanford, CA 94305}

\author{C.~Bartram}
\affiliation{Kavli Institute for Particle Astrophysics and Cosmology, Stanford University, Stanford, CA 94305}
\affiliation{SLAC National Accelerator Laboratory, Menlo Park, CA 94025}

\author{J.~Begin}
\affiliation{Department of Physics, Princeton University, Princeton, NJ 08544}

\author{C.~Bell}
\affiliation{Department of Physics, Stanford University, Stanford, CA 94305}
\affiliation{Kavli Institute for Particle Astrophysics and Cosmology, Stanford University, Stanford, CA 94305}

\author{J.~N.~Benabou}
\affiliation{Berkeley Center for Theoretical Physics, University of California, Berkeley, CA 94720, U.S.A.}
\affiliation{Theoretical Physics Group, Lawrence Berkeley National Laboratory, Berkeley, CA 94720, U.S.A.}

\author{L.~Brouwer}
\affiliation{Accelerator Techology and Applied Physics Division, Lawrence Berkeley National Laboratory, Berkeley, CA 94720}

\author{S.~Chaudhuri}
\affiliation{Department of Physics, Princeton University, Princeton, NJ 08544}

\author{H.~-M.~Cho}
\affiliation{Kavli Institute for Particle Astrophysics and Cosmology, Stanford University, Stanford, CA 94305}
\affiliation{SLAC National Accelerator Laboratory, Menlo Park, CA 94025}

\author{J.~Corbin}
\affiliation{Department of Physics, Stanford University, Stanford, CA 94305}
\affiliation{Kavli Institute for Particle Astrophysics and Cosmology, Stanford University, Stanford, CA 94305}

\author{W.~Craddock}
\affiliation{SLAC National Accelerator Laboratory, Menlo Park, CA 94025}

\author{S.~Cuadra}
\affiliation{Center for Theoretical Physics, Massachusetts Institute of Technology, Cambridge, MA 02139}

\author{A.~Droster}
\email{adroster@berkeley.edu}
\affiliation{Department of Nuclear Engineering, University of California, Berkeley, Berkeley, CA 94720}

\author{J.~Echevers}
\affiliation{Department of Nuclear Engineering, University of California, Berkeley, Berkeley, CA 94720}

\author{J.~W.~Foster}
\email{jwfoster@mit.edu}
\affiliation{Center for Theoretical Physics, Massachusetts Institute of Technology, Cambridge, MA 02139}

\author{J.~T.~Fry}
\affiliation{Laboratory of Nuclear Science, Massachusetts Institute of Technology, Cambridge, MA 02139}

\author{P.~W.~Graham}
\affiliation{Department of Physics, Stanford University, Stanford, CA 94305}

\author{R.~Henning}
\affiliation{Department of Physics and Astronomy, University of North Carolina, Chapel Hill, Chapel Hill, North Carolina, 27599}
\affiliation{Triangle Universities Nuclear Laboratory, Durham, NC 27710}

\author{K.~D.~Irwin}
\affiliation{Department of Physics, Stanford University, Stanford, CA 94305}
\affiliation{Kavli Institute for Particle Astrophysics and Cosmology, Stanford University, Stanford, CA 94305}
\affiliation{SLAC National Accelerator Laboratory, Menlo Park, CA 94025}

\author{F.~Kadribasic}
\affiliation{Department of Physics, Stanford University, Stanford, CA 94305}
\affiliation{Kavli Institute for Particle Astrophysics and Cosmology, Stanford University, Stanford, CA 94305}

\author{Y.~Kahn}
\affiliation{Department of Physics, University of Illinois at Urbana-Champaign, Urbana, IL 61801}

\author{A.~Keller}
\affiliation{Department of Nuclear Engineering, University of California, Berkeley, Berkeley, CA 94720}

\author{R.~Kolevatov}
\affiliation{Department of Physics, Princeton University, Princeton, NJ 08544}

\author{S.~Kuenstner}
\affiliation{Department of Physics, Stanford University, Stanford, CA 94305}
\affiliation{Kavli Institute for Particle Astrophysics and Cosmology, Stanford University, Stanford, CA 94305}

\author{A.~Kunder}
\affiliation{Department of Physics, Stanford University, Stanford, CA 94305}
\affiliation{Kavli Institute for Particle Astrophysics and Cosmology, Stanford University, Stanford, CA 94305}

\author{N.~Kurita}
\affiliation{SLAC National Accelerator Laboratory, Menlo Park, CA 94025}

\author{A.~F.~Leder}
\affiliation{Department of Nuclear Engineering, University of California, Berkeley, Berkeley, CA 94720}
\affiliation{Physics Division, Lawrence Berkeley National Laboratory, Berkeley, CA 94720}

\author{D.~Li}
\affiliation{Kavli Institute for Particle Astrophysics and Cosmology, Stanford University, Stanford, CA 94305}
\affiliation{SLAC National Accelerator Laboratory, Menlo Park, CA 94025}

\author{N.~Otto}
\affiliation{Department of Physics, Princeton University, Princeton, NJ 08544}

\author{J.~L.~Ouellet}
\affiliation{Laboratory of Nuclear Science, Massachusetts Institute of Technology, Cambridge, MA 02139}

\author{K.~M.~W.~Pappas}
\affiliation{Laboratory of Nuclear Science, Massachusetts Institute of Technology, Cambridge, MA 02139}

\author{A.~Phipps}
\affiliation{Department of Physics, California State University, East Bay, Hayward, CA 94542}

\author{N.~M.~Rapidis}
\email{rapidis@stanford.edu}
\affiliation{Department of Physics, Stanford University, Stanford, CA 94305}
\affiliation{Kavli Institute for Particle Astrophysics and Cosmology, Stanford University, Stanford, CA 94305}

\author{B.~R.~Safdi}
\affiliation{Berkeley Center for Theoretical Physics, University of California, Berkeley, CA 94720, U.S.A.}
\affiliation{Theoretical Physics Group, Lawrence Berkeley National Laboratory, Berkeley, CA 94720, U.S.A.}

\author{C.~P.~Salemi}
\affiliation{Department of Physics, Stanford University, Stanford, CA 94305}
\affiliation{Kavli Institute for Particle Astrophysics and Cosmology, Stanford University, Stanford, CA 94305}
\affiliation{SLAC National Accelerator Laboratory, Menlo Park, CA 94025}

\author{M.~Simanovskaia}
\affiliation{Department of Physics, Stanford University, Stanford, CA 94305}
\affiliation{Kavli Institute for Particle Astrophysics and Cosmology, Stanford University, Stanford, CA 94305}

\author{J.~Singh}
\affiliation{Department of Physics, Stanford University, Stanford, CA 94305}
\affiliation{Kavli Institute for Particle Astrophysics and Cosmology, Stanford University, Stanford, CA 94305}

\author{P.~Stark}
\affiliation{Department of Physics, Stanford University, Stanford, CA 94305}
\affiliation{Kavli Institute for Particle Astrophysics and Cosmology, Stanford University, Stanford, CA 94305}
\affiliation{SLAC National Accelerator Laboratory, Menlo Park, CA 94025}

\author{E.~C.~van~Assendelft}
\affiliation{Department of Physics, Stanford University, Stanford, CA 94305}
\affiliation{Kavli Institute for Particle Astrophysics and Cosmology, Stanford University, Stanford, CA 94305}

\author{K.~van~Bibber}
\affiliation{Department of Nuclear Engineering, University of California, Berkeley, Berkeley, CA 94720}

\author{K.~Wells}
\affiliation{Department of Physics, Stanford University, Stanford, CA 94305}

\author{J.~Wiedemann}
\affiliation{Department of Physics, Princeton University, Princeton, NJ 08544}

\author{L.~Winslow}
\affiliation{Laboratory of Nuclear Science, Massachusetts Institute of Technology, Cambridge, MA 02139}

\author{W.~J.~Wisniewski}
\affiliation{SLAC National Accelerator Laboratory, Menlo Park, CA 94025}

\author{D.~Wright}
\affiliation{Kavli Institute for Particle Astrophysics and Cosmology, Stanford University, Stanford, CA 94305}
\affiliation{Department of Electrical Engineering, Stanford University, Stanford, CA 94305}

\author{A.~K.~Yi}
\affiliation{Kavli Institute for Particle Astrophysics and Cosmology, Stanford University, Stanford, CA 94305}
\affiliation{SLAC National Accelerator Laboratory, Menlo Park, CA 94025}

\author{B.~A.~Young}
\affiliation{Department of Physics, Santa Clara University, Santa Clara, CA 95053}

\collaboration{DMRadio Collaboration}
\noaffiliation

\date{\today}

\begin{abstract}

DMRadio-m$^3$ is an experiment that is designed to be sensitive to KSVZ and DFSZ QCD axion models in the 10--200\,MHz (41\,neV$/c^2$ -- 0.83\,$\mu$eV/$c^2$) range. The experiment uses a solenoidal dc magnetic field to convert an axion dark-matter signal to an ac electromagnetic response in a coaxial copper pickup. The current induced by this axion signal is measured by dc SQUIDs. In this work, we present the electromagnetic modeling of the response of the experiment to an axion signal over the full frequency range of DMRadio-m$^3$, which extends from 
the low-frequency, lumped-element limit to a regime where the axion Compton wavelength is only a factor of two larger than the detector size. With these results, we determine the live time and sensitivity of the experiment. The primary science goal of sensitivity to DFSZ axions across 30--200 MHz can be achieved with a $3\sigma$ live scan time of 2.9 years.

\end{abstract}

\maketitle


%
%

\newcommand{\gagg}{\ensuremath{g_{a\gamma\gamma}}\xspace}
\newcommand{\gagggann}{\ensuremath{g_{a\gamma\gamma}g_{aNN}\xspace}}
\newcommand{\gagggaee}{\ensuremath{g_{a\gamma\gamma}g_{aee}\xspace}}

\newcommand{\rhoDM}{\ensuremath{\rho_{\rm DM}}\xspace}
\newcommand{\Jeff}{\ensuremath{\mathbf{J}_{\rm eff}}\xspace}

\newcommand{\cPU}{\ensuremath{c_{\rm PU}}\xspace}
\newcommand{\VPU}{\ensuremath{V_{\rm PU}}\xspace}
\newcommand{\LPU}{\ensuremath{L_{\rm PU}}\xspace}
\newcommand{\Leff}{\ensuremath{L_{\rm eff}}\xspace}

\newcommand{\DMR}{\mbox{DMRadio}\xspace}
\newcommand{\DMRp}{\mbox{\DMR-Pathfinder}\xspace}
\newcommand{\DMRL}{\mbox{\DMR-50L}\xspace}
\newcommand{\DMRm}{\mbox{\DMR-m$^3$}\xspace}
\newcommand{\DMRGUT}{\mbox{\DMR-GUT}\xspace}

\newcommand{\ABRA}{\mbox{ABRACADABRA}\xspace}
\newcommand{\ABRAten}{\mbox{ABRACADABRA-10\,cm}\xspace}


%
%

\begin{acronym}
\acro{SM}{Standard Model}
\acro{QED}{quantum electrodynamics}
\acro{QCD}{quantum chromodynamics}
\acro{BSM}{beyond the standard model}
\acro{DM}{dark matter}
\acro{CDM}{cold dark matter}
\acro{GUT}{grand unification theory}
\acro{WIMP}{weakly interacting massive particle}
\acro{SHM}{Standard Halo Model}
\acro{ppm}{part-per-million}
\acro{ppb}{part-per-billion}

\acro{ADM}{axion dark matter}
\acro{ALP}{axion-like particle}
\acro{PQ}{Peccei-Quinn}
\acro{PQWW}{Peccei-Quinn-Wilczek-Weinberg}
\acro{KSVZ}{Kim-Shifman–Vainshtein–Zakharov}
\acro{DFSZ}{Dine–Fischler–Srednicki–Zhitnitsky}
\acro{LSW}{light shining through wall}

\acro{DP}{dark photon}

\acro{DR}{dilution refrigerator}
\acro{PT}{pulse tube}
\acro{OFHC}{oxygen-free, high-conductivity}

\acro{TE}{transverse electric}
\acro{TM}{transverse magnetic}
\acro{TEM}{transverse electromagnetic}
\acro{MQS}{magneto-quasistatic}
\acro{SQL}{standard quantum limit}
\acro{QND}{quantum non-demolition}

\acro{DFT}{discrete Fourier transform}
\acro{FFT}{fast Fourier transform}
\acro{SNR}{signal-to-noise ratio}
\acro{PSD}{power spectral density}

\end{acronym}

%
\acrodefplural{PSD}{power spectral densities}
\acrodefplural{GUT}{grand unification theories}
\acrodefplural{ppm}{parts-per-million}
\acrodefplural{ppb}{parts-per-billion}


\section{Introduction}
\label{sec:Intro}

The axion is a well-motivated dark matter (DM) candidate that solves the strong charge-parity (CP) problem of quantum chromodynamics (QCD) \cite{Peccei:1977hh,Peccei:1977ur,Weinberg:1977ma,Wilczek:1977pj} while also having a favorable cosmological production mechanism that can populate the universe with the observed cold DM abundance \cite{Abbott:1982af,Preskill1983,Dine1983}. It arises as a pseudo-Goldstone boson of a spontaneously broken Peccei-Quinn symmetry that has a symmetry breaking energy scale $f_a$. Through interactions with QCD, the axion mass is set by the inverse of the symmetry breaking scale as $m_a\approx5.7\text{ neV}(10^{15}\text{ GeV}/f_a)$ \cite{Borsanyi:2016ksw}. Recent theoretical work suggests that the value of $f_a$ could be as high as the Planck scale, which motivates searches that are sensitive to axions with mass in the range of peV$/c^2$--$\mu$eV$/c^2$ \cite{Tegmark:2005dy,Hertzberg:2008wr,Co:2016xti,Graham:2018jyp,takahashi2018qcd}. Such axion and axion-like particle (ALP) models are of interest for grand unification theories (GUTs) \cite{Co:2016xti, DiLuzio:2020wdo, Wise:1981ry, Ballesteros:2016xej, Ernst:2018bib, DiLuzio:2018gqe, Ernst:2018rod,FileviezPerez:2019fku, FileviezPerez:2019ssf}, string theory models \cite{Svrcek:2006yi,Green:1984sg,Conlon:2006tq,Acharya:2010zx,Ringwald:2012cu,Cicoli:2012sz,Halverson:2019cmy,Witten:1984dg}, and naturalness arguments \cite{PhysRevD.73.023505,Graham:2018jyp}.

One of the most sensitive techniques to search for axions takes advantage of their coupling to standard-model photons. An axion in a dc magnetic field can convert to a photon whose frequency matches the axion mass ($\nu_a=m_ac^2/h$). Due to their low mass and low temperature, dark matter axions have a high per-state occupation number and thus behave like a classical field. The axion-photon interaction can then be considered as a modification to Maxwell's equations, resulting in a new effective current that depends on the local magnetic field and the local axion density. Neglecting spatial gradients in the dark matter axion field, which tend to be small, we write the effects of the axion-photon coupling as an effective current in the presence of a dc magnetic field \cite{Sikivie1983}:
\begin{equation}
    \mathbf{J}_\text{eff}= g_{a\gamma\gamma}\frac{\sqrt{\hbar c}}{\mu_0}\sqrt{2\rho_\text{DM}}\cos\left(\frac{m_ac^2}{h}t\right)\mathbf{B}.
    \label{eq:AxCurrentDens}
\end{equation}
Here $g_{a\gamma\gamma}$ is the axion-photon coupling constant, $\rho_{\text{DM}}=0.45\text{ GeV}\text{/cm}^{3}$ is the local axion dark matter density \cite{de_Salas_2021}, $m_a$ is the axion mass, and $\mathbf{B}$ is a background dc magnetic field. Axion-dark-matter experiments typically scan through a frequency band to search for axions down to a specific coupling strength $g_{a\gamma\gamma}$ at a specified signal-to-noise ratio. The metric of sensitivity is given by the scan rate, which denotes the frequency range covered per unit time to achieve sensitivity to axions at a given $g_{a\gamma\gamma}$.

To date, the most sensitive limits have been set by cavity haloscopes in mass ranges above 1\,$\mu$eV$/c^2$ (240 MHz) \cite{PhysRevLett.127.261803,Backes2021,Yi:2022fmn}. When a microwave cavity is placed in a dc magnetic field, axions  produce an effective current that runs along the magnetic field lines  through the cavity. If the frequency of a cavity mode that couples to this current matches that of the axion, the signal is resonantly enhanced. 
Such cavity experiments become impractical at low axion frequencies due to the size of the cavities that is required to support cavity modes with long wavelength.

Experiments in the sub-$\mu$eV mass range decouple the resonance conditions from the boundary conditions imposed by the physical dimensions of a simple cavity, making it possible to access lower frequencies within practical experimental volumes. At very low frequencies,  these detectors are often referred to as ``lumped element,'' because they isolate electric and magnetic field energy into capacitive and inductive elements \cite{Cabrera2008,sikivie2014proposal}. Experiments in operation or proposed in the lumped-element limit include ADMX-SLIC, a solenoidal experiment \cite{Crisosto:2019fcj}, as well as toroidal experiments such as the \ABRAten prototype \cite{PhysRevLett.122.121802,PhysRevD.99.052012,PhysRevLett.127.081801,PhysRevLett.117.141801}, SHAFT \cite{Gramolin2020a}, and the DMRadio-50L experiment \cite{Rapidis:2022proceedings}.
\DMRp performed a resonant search for \acp{DP} in the lumped-element regime with a solenoidal pickup \cite{10.1007/978-3-030-43761-9_16}.  A toroidal
geometry is also analyzed in the regime ranging from the lumped-element limit to higher frequencies \cite{Benabou:2022arx}. Other ways to probe lower frequency axions in a subwavelength space include nuclear spin coupling \cite{Budker:2013hfa} and heterodyne frequency conversion \cite{PhysRevLett.123.021801}.

In this paper, we model the sensitivity of the \DMRm{} experiment \cite{Brouwer:2022DMRm}, which operates a coaxial pickup in a $\sim4.7$\,T solenoidal magnet at frequencies both within and above the lumped-element limit, where lumped element approximations are not accurate \cite{Benabou:2022arx}. Here, we analyze the sensitivity by modeling the experiment as an equivalent tuned series resistor-inductor-capacitor (RLC) circuit with numerical results from Finite Element Modeling (FEM) simulations. Incorporating realistic noise models for dc SQUIDs, we  use the formalism in \cite{Chaudhuri:2018rqn} to determine the scan times required for each frequency interval. We present a baseline design for \DMRm{} to achieve sensitivity to Dine-Fischler-Srednicki-Zhitnisky (DFSZ) \cite{Dine:1981rt, Zhitnitsky:1980tq} axions in the 30--200\,MHz  range as the primary science goal of the experiment, as well as sensitivity to Kim-Shifman-Vainshtein-Zakharov (KSVZ) \cite{PhysRevLett.43.103,SHIFMAN1980493} axions in the 10--30\,MHz range, as the secondary science goal of the experiment. We also present a plan to extend the search to axions with a coupling of $g_{a\gamma\gamma}=g_{a\gamma\gamma,\text{DFSZ}}(\nu=30\text{ MHz})=1.87\times10^{-17}\text{ GeV}^{-1}$ across 5--30 MHz, thus extending the parameter space  covered in the secondary science goal. We denote this as the extended goal.

In $\S$\ref{sec:lumped_element_detectors} we motivate and describe the coaxial design of the \DMRm{} pickup structure and present the constraints imposed by the magnet.  In  $\S$\ref{sec:OpenCircuitEigenmodes}, we start with an intuitive understanding of the mode structure of the coaxial pickups. We then discuss the different coaxial pickups used for different frequency ranges, and summarize their computed eigenmodes. We show that \DMRm{} uses a single TEM-like mode, well-separated from other parasitic modes, over the science bandwidth of each coaxial element. This approach enables the formalism of \cite{Chaudhuri:2018rqn} to be rigorously applied.
 In $\S$\ref{sec:equivalent_RLC}, we numerically extract an equivalent series RLC circuit representing the scanned resonance, which is then used to compute the science reach of \DMRm{}. In $\S$\ref{sec:tuning}, we discuss the tuning requirements and constraints for the experiment. $\S$\ref{sec:squids} presents relevant details on the SQUID readout amplifiers as they apply to \DMRm{}. Finally, in $\S$\ref{sec:ScienceReach}, we determine the  required $3\sigma$ live scan times for \DMRm{} to achieve its science goals.

\section{\DMRm pickup geometry}
\label{sec:lumped_element_detectors}

\begin{figure*}[ht]
    \centering
    \includegraphics[width=.99\textwidth]{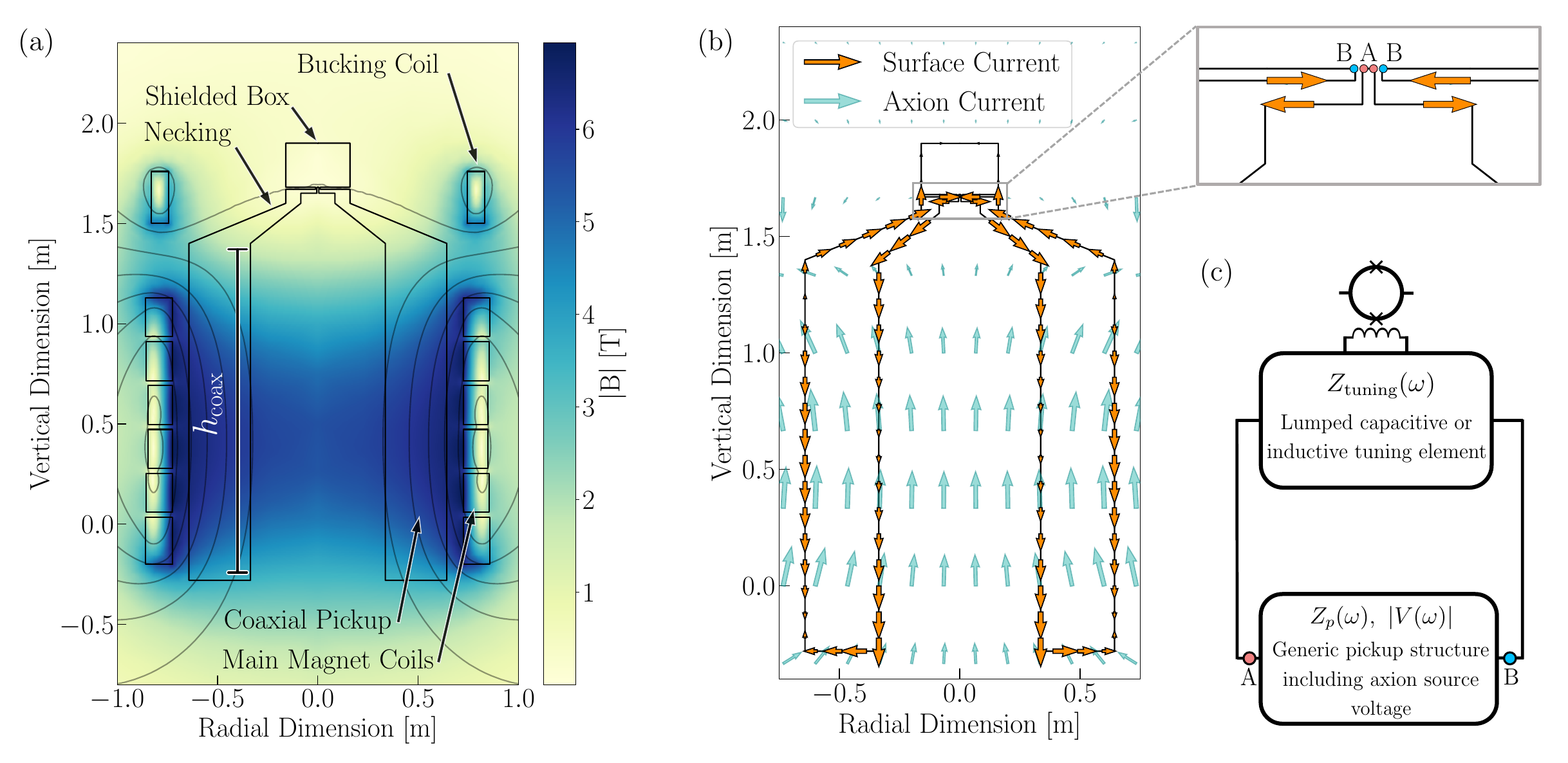}
    \caption{(a) A 2D cross section of the dc magnetic field profile of \DMRm produced by the main magnet coil and the bucking coils, as well as the full coaxial pickup cross section. These structures exist in a shielded vacuum space whose boundary is not shown. In this figure, the coax height, $h_\text{coax}$, is $1.68$ m. The main magnet coils produce a $B_0\approx 5$\,T region for coupling to axion signals, and the shielded box encloses a low-magnetic-field region at the top. The low-field region has additional magnetic and EMI shielding, and contains superconducting components (not shown) that include the tuning element and the SQUID readout. (b) A cross-section of the coaxial pickup showing the effective axion current (light-green arrows) and the resulting induced surface current due to screening effects on the inner surface of the coaxial pickup (orange arrows). This simulation was run at 100 MHz. The nodes in the surface current can be understood as an effect of \textit{bulk electron shuttling} (see Sections \ref{sec:equivalent_RLC} and Appendix \ref{sec:AppShuttling} for details).  The voltage and impedance of the circuit are measured across the two rings defined by $\text{A}$ and $\text{B}$ respectively, as shown in the expanded view. (c) The circuit parameters of this pickup structure are measured as a frequency-dependent reactance $Z_p(\omega)$ and axion-signal voltage amplitude $|V(\omega)|$ measured between the circles marked A and B in subfigure (b). This impedance is modeled as a series RLC circuit ($\S$ \ref{sec:equivalent_RLC}) and is coupled to the impedance of a reactive tuning element that is located in the shielded box. At each frequency, the impedance parameters, the voltage amplitude induced by the axion signal, and the amplifier properties determine the scan rate.
    }
    \label{fig:circmodel}
\end{figure*}

The pickup geometry of \DMRm{} is coaxial. For a coaxial transmission line, transverse electromagnetic (TEM) propagating waves have no lower cutoff frequency.
A coaxial transmission line resonator is thus a natural geometry for use in axion experiments at wavelengths larger than the length scale of the experiment 
providing access to low frequencies through a tunable resonant mode based on a TEM propagating mode in the coaxial section.  As shown schematically in Figure \ref{fig:circmodel}a, \DMRm{} consists of a copper coaxial pickup structure (sometimes referred to as the coax) in the region of highest magnetic field in the bore of a solenoidal magnet. The straight coaxial section forms the majority of the pickup, with a fixed bottom plate shorting the bottom. The top end of the pickup necks down to a superconducting tuning circuit in a small low-field shielded region enclosed in a pillbox-shaped container (henceforth referred to as the shielded box). This shielded box holds the reactive tuning elements and SQUID readout amplifiers. Figure \ref{fig:circmodel}a shows a cross-section of one coaxial pickup without any tuning or readout elements. 

The tuning and readout elements for \DMRm{} are connected across the rings that are denoted as points A and B in Figure \ref{fig:circmodel}b and are placed in the shielded box that encloses the low-field region shown in Figure \ref{fig:circmodel}a. The electromagnetic properties of the coax can be modeled and understood independently from the tuning and readout components, as shown in the circuit model in Figure \ref{fig:circmodel}c. 


The dc magnetic field strength and volume for \DMRm{} are optimized to maximize science reach for a given magnet cost. The chosen magnet dimensions set a maximum coaxial outer radius of $r_{o}=0.628$\,m. The coaxial inner radius is chosen to be $r_i=0.336$\,m. Six interchangeable copper coaxial pickups with different values of $h_{\rm coax}$ (defined in Figure \ref{fig:circmodel}a) are used, the tallest of which has a height, $h_\text{coax}= 1.68$\,m. Different coaxial lengths are optimal for use at different frequencies. For \DMRm{}, the lowest axion frequency of interest (5\,MHz) corresponds to a Compton wavelength of 60\,m; the highest axion frequency of interest (200\,MHz) corresponds to Compton wavelength 1.5\,m. Thus, \DMRm{} spans frequencies well above the lumped-element limit. In the following section ($\S$\ref{sec:OpenCircuitEigenmodes}), we discuss the coax in terms of its cavity properties; in $\S$\ref{sec:equivalent_RLC} we show how to describe the coax as an equivalent circuit at each frequency, and then how to determine its sensitivity.

\section{Modes of the \DMRm coax}
\label{sec:OpenCircuitEigenmodes}

\DMRm{} utilizes a set of six coaxial pickups to achieve the science goals of the experiment. The underlying physical motivation behind the use of six coaxial pickups is three-fold:
\begin{enumerate}[(a)] 
\item The resonator is designed to be single-moded at each frequency that it scans. In order to achieve this, the coax is operated below the frequency at which a second mode appears.
\item For a given length of a coaxial pickup, the impedance of the structure and axion-induced voltage vary with frequency; these quantities in turn influence the scan rate. The use of different sized pickups provides an acceptable scan rate at all frequencies of interest.
\item The tuning elements required at each frequency must be physically realizable, and must enable tuning over a wide frequency range. 
\end{enumerate}
Point (a) is discussed in this section, while points (b) and (c) are discussed in Sections  \S \ref{sec:equivalent_RLC} and \S\ref{sec:tuning} respectively.

We first provide a simplified intuitive model to understand the resonant features of the \DMRm{} coax. Subsequently we describe the eigenmodes of the structure and how they influence the scan strategy of the experiment.

\subsection{A coaxial toy model}
\label{sec:subtoymodel}
To qualitatively understand \DMRm{}, we consider the model of an idealized lossless coaxial structure of uniform outer radius $r_o$, inner radius $r_i$, and height $h_{\rm coax}$, shorted at the bottom and open at the top. The axion signal due to the solenoidal magnetic field generates a voltage between the top of the outer and inner cylinders of the coax. The impedance of the coaxial pickup seen at its open top end is given by the equation for the impedance of a lossless coaxial cable shorted at one end \cite{pozar2011microwave}:
\begin{equation}
Z_p(\omega)=iX_p(\omega)=i Z_c \tan\left(\frac{\omega h_{\rm coax}}{c}\right),
\label{eq:Zcoax}
\end{equation}
where $X_p$ is the reactance of $Z_p$. The characteristic impedance of the coax is:
\begin{equation}
Z_c = \frac{Z_0}{2\pi}\ln\left(\frac{r_o}{r_i}\right),
\label{eq:Zchar}
\end{equation}
where $Z_0\approx 377\,\Omega$ is the impedance of free space. If left open, this coax has TEM transmission line resonances at $\nu=(2n-1) c/(4h_\text{coax})$, for $n=1,2,3...$, the first being the  quarter wavelength ($\lambda/4$) mode. 
If shorted, this coax has TEM transmission line resonances starting at the half wavelength mode ($\lambda/2$) where $\nu=c/(2h_\text{coax})$. 

More generally, the resonance frequency of the coaxial transmission line resonator of Eq. \ref{eq:Zcoax} can be adjusted by loading it with a reactance, $X_\text{tuning}.$ At a target resonance frequency, $\omega_0$, the tuning reactance is chosen to be equal and opposite to the coaxial pickup reactance, $X_\text{tuning}(\omega_0)=-X_p(\omega_0)$, thus satisfying the condition for resonance where the total reactance is zero. Depending on the sign of $X_p(\omega_0)$, the tuning element may either be inductive or capacitive.

In the lower frequency ranges of \DMRm{}, the coaxial impedance is well approximated as a linear inductor with $Z_\text{coax}\approx i\omega L_\text{coax}= i \omega Z_c h_{\rm coax}/c$. This limit is the lumped-element limit, in which the voltage across the output is proportional to the time derivative of the coupled magnetic flux from the axion signal, since signals propagate with negligible phase lag from one end of the structure to the other. A tuning capacitance is required to achieve the resonant condition. At higher frequencies, the reactance is no longer linear with frequency, and when 
$\pi/2<\omega_0 h_{\rm coax}/c<\pi$, $Z_p$ is capacitive, and a tuning inductance is required to achieve resonance.

\subsection{Eigenmodes of the \DMRm{} coax}


In this paper we use the formalism developed in \cite{Chaudhuri:2018rqn} to compute the scan rates required to achieve our science goals. As that formalism applies to a single-moded resonator and since coaxial structures can support multiple cavity modes, the eigenmodes of the \DMRm{} pickup must be simulated and their characteristics understood. \DMRm{} is designed to avoid mode degeneracies.

As discussed in the previous section, the \DMRm{} coaxes can be understood by considering a simple coaxial cavity with one end shorted and the other end reactively loaded. In the case where both ends are shorted, resonance frequencies of all transverse electric (TE), transverse magnetic (TM), and TEM modes can be calculated from closed-form formulas for uniform coaxes, and with finite-element modeling for realistic detectors, thus providing a baseline for our understanding of the mode structure.

The TM modes of \DMRm{} coaxes have resonance frequencies above 200 MHz and do not influence the scan strategy in any way. However, the lowest order TE mode, $\text{TE}_{111}$, occurs at a lower frequency, raising a concern of mode degeneracy. The resonance frequency of the $\text{TE}_{111}$ mode for an idealized coax with length much larger than either radial dimension can be approximated as:
\begin{equation}
    \nu_{\text{TE}_{111}}\approx \frac{2c}{2\pi r_i \left(1+\frac{r_o}{r_i}\right)},
\end{equation}
where $r_i$ and $r_o$ are the inner and outer radii of the coax, respectively \cite{pozar2011microwave}. In this long coax limit, the values of $r_{o}=0.628$\,m and $r_{i}=0.336$\,m for \DMRm{} support a TE$_{111}$ mode with $\nu_{\text{TE}_{111}}\approx 100$\,MHz. Shorter coaxes will have higher $\nu_{\text{TE}_{111}}$.
In the case of \DMRm{}, where the coax is shorted at one end and tapers at the other end, these modes are neither analytically solvable nor do they have a trivial field profile. We determine the precise values using the FEM  software \texttt{COMSOL} \cite{comsol}, which confirms that $\nu_{\text{TE}_{111}}$ is above 100\,MHz for all \DMRm{} coaxes, and that shorter coaxes have higher $\nu_{\text{TE}_{111}}$. In this section we show that careful choice of coaxes places  $\nu_{\text{TE}_{111}}$ above all operational frequencies, and with $\nu_{\text{TE}_{111}}$ at 200 MHz for the shortest coax of $h_\text{coax}=0.56$\,m.

All modes in such a structure still form a complete set of eigenstates and, in an ideal case, the TE-like modes would not mix with the TEM-like mode that is being used for an axion scan. 
This idealized approximation where the two modes do not mix is broken, in part, by the readout coupling mechanism that in practice will not have pure azimuthal symmetry. As a result, TE-like modes have a non-zero mixing with the TEM-like mode. We design all coaxial elements to operate only at frequencies below the TE$_{111}$ mode.

\begin{figure}
    \centering
    \includegraphics[width=.49\textwidth]{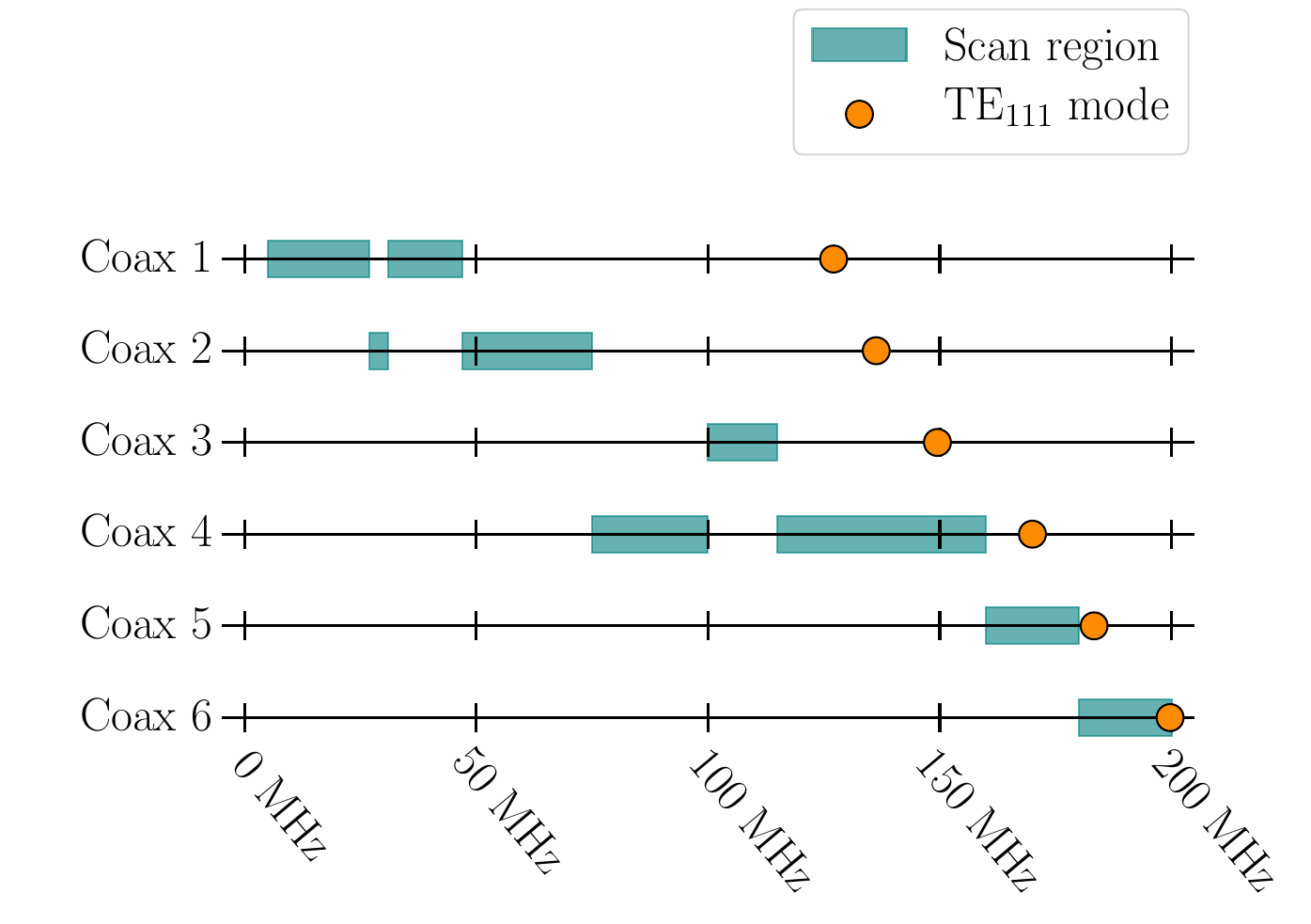}
    \caption{Graphical representation of the frequency range of scan regions (Table \ref{tab:CoaxChoices}) and the corresponding TE$_{111}$-like modes. \DMRm{} is designed to avoid the TE$_{111}$-like modes due to potential mode mixing with the TEM-like mode that is used for the axion scan. Simulations have shown that the frequency of the TE-like mode does not significantly shift from loading by the tuning elements, and it is possible to stay at frequencies below this mode in operation.}
    \label{fig:modes}
\end{figure}

Informed by the mode structure of these coaxial pickups, as well as the results from the next section in which we use \texttt{COMSOL} to extract the scan rate of the experiment, we select a set of six coaxial pickups to scan the \DMRm{} science region. These coaxial pickups, frequency ranges, and their respective TE$_{111}$ mode resonances are presented in Table \ref{tab:CoaxChoices} and  Figure \ref{fig:modes}.  These simulations  also show that the resonant frequencies of TE-like modes shift by less than $1\%$ when a reactive load is placed across points A and B (Figure \ref{fig:circmodel}b) of the coaxial pickup relative to when it is open. This means that even as the TEM-like scanning mode is tuned across a range of frequencies, the TE-like mode's frequency remains sufficiently stable to be avoided.

\begin{table}[]
    \centering
\begin{tabular}{c|c|c}
Coax Number & $h_\text{coax}$ & Frequency Range(s) (MHz)                                             \\ \hline
1           & 1.68\,m     &\begin{tabular}[c]{@{}c@{}}5--27\,MHz\\ 31--47\,MHz\end{tabular}                                                          \\ \hline
2         & 1.40\,m  & \begin{tabular}[c]{@{}c@{}}27--31\,MHz\\ 47--75\,MHz\end{tabular}                                                    \\ \hline
3           & 1.12\,m     & 100--115\,MHz                                                        \\ \hline
4           & 0.84\,m     & \begin{tabular}[c]{@{}c@{}}75--100\,MHz\\ 115--160\,MHz\end{tabular} \\ \hline
5           & 0.70\,m      & 160--180\,MHz                                                        \\ \hline
6           & 0.56\,m     & 180--200\,MHz                                                       
\end{tabular}
    \caption{Table of the six coaxial pickups (coaxes) to be used in \DMRm for the primary and secondary science goals as well as the  extended  goal. Together, these coaxes enable frequency scans throughout the 5--200 MHz range. The quantity $h_\text{coax}$ is shown in Figure \ref{fig:circmodel}a. The frequency range chosen for each coaxial pickup optimizes simultaneously the constraints of practical tuning elements and the scan rate, while requiring that the resonator be single-moded. Some coaxes cover two discrete frequency ranges. These conditions are further discussed in \S\ref{sec:equivalent_RLC}.  
}
    \label{tab:CoaxChoices}
\end{table}

For Coaxes 1-5 listed in Table \ref{tab:CoaxChoices}, the TE$_{111}$ mode is found in the $125\text{ MHz}<f<185\text{ MHz}$ range. However, by judiciously choosing the scan range of each coax, it can be avoided in each case, as shown in Figure \ref{fig:modes}. In Coax 6, the TE$_{111}$ mode exists at 200 MHz and, as such, the end of the single-moded frequency range for the shortest coax coincides with the upper boundary of the \DMRm{} science range.

As is discussed in more detail in Appendix \ref{sec:Box}, the shielded box at the top of the coax is  small enough that it has no cavity modes within the signal band, and is large enough that it does not significantly capacitively shunt the coaxial pickup. Numerical models indicate that, apart from these two extremes, the performance of the detector is not significantly affected by the dimensions of the shielded box. Radiation to free space is not a consideration because the experiment is fully shielded.

From the cavity analysis above, we conclude that \DMRm{} always operates in a single-moded regime, where the formalism from \cite{Chaudhuri:2018rqn} applies. In the next section, we explain how the formalism is implemented in our sensitivity analysis.

\section{The Equivalent series RLC resonator}
\label{sec:equivalent_RLC}

\DMRm{} uses a coaxial pickup similar to the toy model discussed in \S\ref{sec:subtoymodel}, but with additional features including a neckdown region to a shielded box that contains the tuning reactances and dc SQUID amplifiers. This design results in an impedance that is qualitatively similar to that described by Eq. \ref{eq:Zcoax}, but that requires numerical computation with modeling software described in the following section. The voltage signal induced by axions is similarly difficult to compute in closed form and is numerically computed. 
It is worth noting that in typical cavity haloscopes, the axion does not couple to TE and TEM modes \cite{ADMX:2019uok,Brubaker2017,Rapidis:2018dci} as explained using the form factor formalism \cite{Sikivie1983}.
As is shown in this paper, the \DMRm{} TEM-like resonance does indeed couple both in the quasistatic limit and at frequencies beyond where this approximation applies.


Having established in the previous section that \DMRm{} has a single tunable TEM-like mode at its operational frequencies, we now compute the scan rate by a consideration of the mode impedance and the signal amplitude. From \cite{Chaudhuri:2018rqn}, we write the scan rate equation in the form that we use in this study: 
\begin{equation}
\begin{split}
\frac{d\nu}{dt}&=\frac{\pi(6.4\times10^5)}{\text{SNR}^2}\frac{\hbar^2}{16c^8m_a^4}\times\\
&\qquad\frac{|V(m_a,\mathbf{B}(\mathbf{r}),g_{a\gamma\gamma})|^4Q(\nu_r)\bar{\mathcal{G}}[\nu_r,T,\eta(\nu_r)]}{L_\text{eff}(\nu_r)^2}.
\end{split}
\label{eq:scanrate}
\end{equation}
The numeric pre-factor depends on the physics of a standard halo model \cite{Brouwer:2022DMRm,Herzog-Arbeitman:2017fte}, $|V(m_a,B,g_{a\gamma\gamma})|$ is the axion-mass-dependent magnitude of the induced voltage, which depends on the magnetic field strength and axion-photon coupling, and $Q(\nu_r)$ is the quality factor of the resonator at the resonator frequency $\nu_r$. $\bar{\mathcal{G}}[\nu_r,T,\eta(\nu_r)]$, which is evaluated at $T=0.02$\,K for \DMRm{}, parametrizes noise physics and is discussed in $\S$\ref{sec:squids}, as well as in Appendix F4 of \cite{Chaudhuri:2018rqn} and in Eqs. (A12) and (A13) of \cite{Brouwer:2022DMRm}. $L_\text{eff}$ is the total effective inductance of the combined equivalent RLC circuit, and SNR is the signal to noise ratio. We set $\text{SNR}=3$, which corresponds to a $3\sigma$ axion signal. 

In our simulations presented this section, we use the finite-element modeling package \texttt{COMSOL} to numerically extract the impedance $Z_p(\omega)$ and the voltage induced by the axion $V(\omega)$ across the coaxial pickup. We then model $Z_p(\omega)$ at each frequency as a series RLC circuit.  Subsequently, combining this equivalent circuit with an equivalent circuit of the tuning elements, we extract $L_\text{eff}(\nu_r)$, $Q(\nu_r)$, and $|V(m_a,B,g_{a\gamma\gamma})|$ for the computation of scan rate in Eq. \ref{eq:scanrate}.


\subsection{Extracting the effective RLC circuit}
\label{sec:RLC}

\begin{figure}[ht]
    \centering
    \includegraphics[width=.48\textwidth]{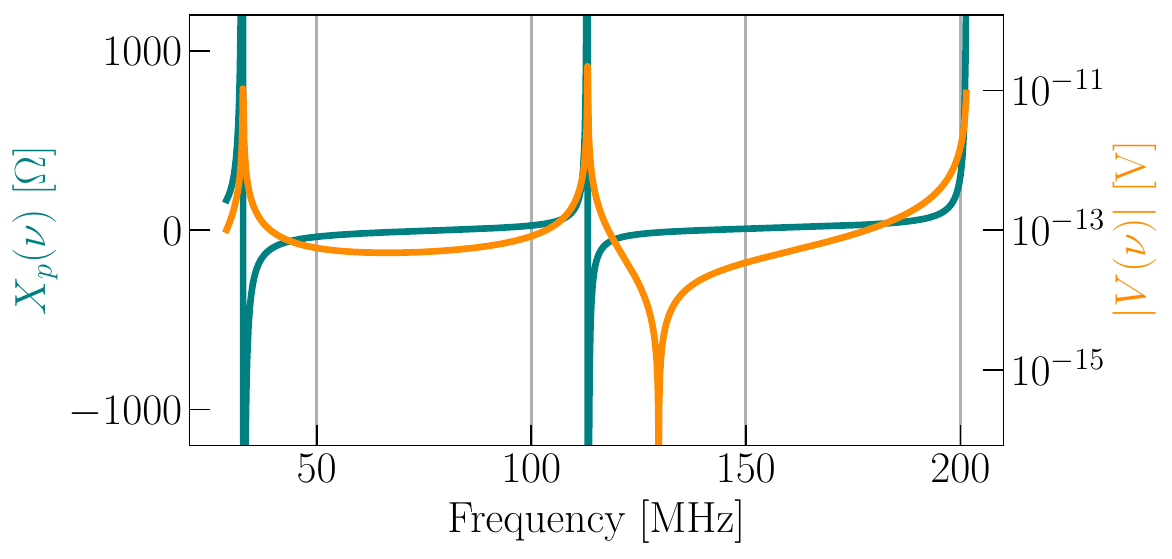}
    \caption{ The reactance, $X_p(\nu)=\text{Im}(Z_p(\nu))$. numerically extracted across $30-$200 MHz  for the $h_\text{coax}=1.4$\,m coaxial pickup (green) and the corresponding voltage induced by a DFSZ axion with the local dark matter density of $\rho_\text{DM}=0.45\text{ GeV/cm}^3$ (orange) \cite{de_Salas_2021}. This coax length corresponds to Coax 2 in Table \ref{tab:CoaxChoices} and is plotted here as an example. The impedance of the tuning elements is not included; these values are for an open, untuned structure. The precise modeled reactance $X_p(\nu)$ is in qualitative agreement with the reactance of a propagating TEM mode in an idealized uniform lossless coax shorted at one end (Eq. \ref{eq:Zcoax}). The zeros 
    and the poles
    of the impedance function represent resonant modes of the untuned coaxial structure. These are TEM-like modes. The resistive portion $R_p(\nu)=\text{Re}(Z_p(\nu))$, not shown in this plot, also has peaks at the mode frequencies. At each point along the $X_p(\nu)$ and $R_g(\nu)$ curves, we Laurent expand to extract an effective series RLC, which can be used to determine the experimental sensitivity. The tuning component is capacitive when $X_p(\nu_r)>0$ and inductive when $X_p(\nu_r)<0$. The voltage curve, $|V(\nu)|$, drops to prohibitively low values in certain frequency regions ($\sim$120--130\,MHz in this untuned structure). This influences the scan rate which scales as $|V(\nu)|^4$ and, thus, such frequency regions are to be avoided.
    }
    \label{fig:impandvolt}
\end{figure}

Using \texttt{COMSOL}, we determine both $V(\omega)$ and $Z_p(\omega)$ between 5 MHz and 200 MHz at $0.1$ MHz frequency intervals for the coaxial pickups without tuning elements. $V(\omega)$ is calculated by simulating the axion signal inside the coaxial pickup. Using a cubic-spline interpolation to fit the real (resistive) and imaginary (reactive) parts of the impedance near any given frequency of interest $\omega_0$, we extract a function for the impedance, $Z_p(\omega)=R_p(\omega)+iX_p(\omega)$, and the derivatives of the resistance and reactance. 
As an example, Figure \ref{fig:impandvolt} shows $X_p(\omega)$ extracted for the \DMRm Coax 2, with $h_{\rm coax}=1.4$\,m. A detailed discussion of the \texttt{COMSOL} and FEM techniques used can be found in Appendix \ref{sec:AppCOMSOL}.

Any impedance function $Z_p(\omega)$ can be represented as a Laurent expansion around the frequency $\omega_0$, for small fractional frequency deviation ($|\omega-\omega_0|\ll\omega_0$), and away from any divergences (such as those caused by internal resonances). Since the frequency-derivative of the
reactance is positive, this expansion can be fully described as a series RLC circuit near $\omega_0$. The derivative of the reactance is always positive in \DMRm{}, and the experiment is designed to operate away from internal resonances.  Using the impedance of a series RLC circuit: $Z_\text{RLC}(\omega)=R+i\omega L-\frac{i}{\omega C}$, as well as its derivative, the parameters of the series RLC equivalent circuit can be determined, forming a new basis set. The two real parts of the series RLC (the resistance and its derivative) map onto the real parts of $Z_p(\omega)$, and the reactive parts have a non-trivial mapping:

\begin{equation}
\begin{aligned}
  R(\omega_0)=&R_p(\omega_0)\\
  \frac{dR}{d\omega}(\omega_0)=& \frac{dR_p}{d\omega}(\omega_0)\\
  L(\omega_0)=&\frac{\omega_0\frac{dX_p}{d\omega}(\omega_0)+X_p(\omega_0)}{2\omega_0}\\
  C(\omega_0)=&\frac{2}{\omega_0\left(\omega_0\frac{dX_p}{d\omega}(\omega_0)-X_p(\omega_0)\right)}.
\end{aligned}
\label{eq:circparams}
\end{equation}
If an effective parallel RLC circuit were used instead, these mappings would have a different form.

We find that at operational frequencies, $\frac{dR_p}{d\omega} \Delta \omega \ll R_p$, where 
$\Delta \omega$ is the sensitivity bandwidth \cite{Chaudhuri:2018rqn,DMRadio:2022jfv}, 
so the derivative of the equivalent resistance $\frac{dR}{d\omega}$ can be taken as zero in the equivalent circuit. We then replace $Z_p(\omega)$ in Figure \ref{fig:circmodel}c for $\omega$ near $\omega_0$ with the series RLC equivalent-circuit impedance
\begin{equation}
    Z_p(\omega)\approx R(\omega_0)+i\omega L(\omega_0) -\frac{i}{\omega C(\omega_0)}.
\end{equation} 
In order to achieve a resonance at frequency $\omega_0$, we introduce a tuning element $Z_\text{tuning}(\omega)$ such that $X_\text{tuning}(\omega_0) = -X_p(\omega_0)$. The tuning element may have additional parasitic loss. The coupling inductor for the dc SQUID imposes additional tuning inductance, which we include as part of $L_\text{tuning}$. 
A similar Laurent expansion can be used for the tuning elements to extract $L_\text{tuning}$, $C_\text{tuning}$, and $R_\text{tuning}$. We define $L_\text{eff} = L + L_\text{tuning}$, $1/C_\text{eff} = 1/C+1/C_\text{tuning}$, $R_\text{eff} = R + R_\text{tuning}$ such that this total effective circuit is our resonant series RLC, with $\omega_0 = 1/\sqrt{L_\text{eff}C_\text{eff}}$ and where $L_\text{eff}(\nu_r)$ enters into Equation \ref{eq:scanrate}. 

\subsection{Calculating the quality factor}
\label{sec:subExtractingQ}

Once the pickup circuit is modeled as a tunable series RLC and the impedance of the tuning and readout components are considered, the quality factor can be determined. The quality factor is given by $Q=\omega_0 L_\text{eff}/R_\text{eff}$, where $L_\text{eff}$ and $R_\text{eff}$ are described in the previous paragraph. 

Since the tuning elements are superconducting but the coaxial pickups are not, $R_\text{eff}$ tends to be dominated by $R_p$. Any dielectric losses in the system are taken to be small as compared to $R_\text{eff}$. In order to compute $R_p$, the \texttt{COMSOL} model starts by modeling the conductivity of room temperature (RT) copper, with no magnetoresistance. To compute the higher ac conductance achieved at cryogenic temperatures, and to take into account magnetoresistance in the dc magnetic field, we increase the conductivity provided by the FEM model by a factor of $6$, resulting in a factor of 6 enhancement in the $Q(\nu)$. The basis of this enhancement factor is discussed in detail in Appendix \ref{sec:AppRRR}.

\subsection{Calculating the axion-induced signal voltage }
We now evaluate the axion-induced voltage across the pickup gap defined by points A and B in Figure \ref{fig:circmodel}b. The effective axion current can be simulated as a source  current proportional to the dc magnetic field strength and following the flow of the magnetic field lines.
By using the ``external current density'' functionality in the \texttt{COMSOL} RF Module, we model the effective axion current that induces an electromagnetic response in the coaxial pickup (further details can be found in Appendix \ref{sec:AppCOMSOL}). The magnetic field used here is numerically modeled in separate simulations. It is important to note that when dc magnetic field lines intersect a metal, the effective axion current also penetrates and traverses the bulk of the metal, as implied by Eq. \ref{eq:AxCurrentDens}. As a consequence of Amp\`ere's law within the bulk of the metal, a physical electron current density flows in the opposite direction in the bulk of the copper, following the magnetic flux lines. It is interesting that this physical electron current penetrates deeper than the skin depth, as a rigorous consequence of modified Maxwell's equations in the presence of axions. This phenomenon, which we refer to as \textit{bulk electron shuttling} (BES), is derived and discussed in more detail in Appendix \ref{sec:AppShuttling}. We find that the \texttt{COMSOL} external current density function accurately models the effective axion current, including bulk electron shuttling. Thus, \texttt{COMSOL} can be used to extract the axion-induced voltage  
 across the pickup. The result of one such voltage simulation is shown in Figure \ref{fig:impandvolt}. This result for $|V(m_a,B,g_{a\gamma\gamma})|$ can then be applied to Equation \ref{eq:scanrate}.

\section{Reactive Tuning Considerations}
\label{sec:tuning}


As discussed in previous sections, the tuning element needed at a given frequency may either be capacitive or inductive, depending on the sign of the coaxial pickup reactance, $X_p(\omega)$, at that frequency. Since the  value of $X_p(\omega)$ may vary significantly over a small frequency range, care must be taken in selecting both the coaxial pickup and the tuning elements to minimize the number of times these elements must be swapped out. The tuning element design is further constrained by the requirement that the tuning elements must fit in the shielded box of the coaxial pickup. 

%



The tuning elements and the connections between the pickup, tuner, and amplifier all present some amount of inductance and capacitance that do not contribute to the sensitivity of the experiment.  These parasitic reactances are carefully controlled so that they do not significantly degrade the experiment's sensitivity. A detailed description of the tuning mechanisms and their corresponding parasitics will be presented in a future work.

\section{SQUID amplifiers in \DMRm{}}
\label{sec:squids}

Finally, we consider the noise behavior of the dc SQUID amplifiers, which determine the parameter $\eta$ in $\bar{\mathcal{G}}[\nu_r,T, \eta(\nu_r)]$ of Equation \ref{eq:scanrate}. Further details on the dc SQUIDs used by \DMRm{} can be found in \cite{DMRmSquids}.

\DMRm{} uses a SQUID-amplifier chain consisting of a variable transformer, a voltage-biased first-stage single SQUID, and a second-stage SQUID array with $50\ \Omega$ output impedance that couples to a room-temperature preamplifier with $50\ \Omega$ input impedance. The first-stage SQUID design is the same at all frequencies. However, two different versions of the second-stage SQUID array are required for different frequency ranges.

The input coil of the first stage dc SQUID is wired in series between the output of the coaxial pickup and the tuning reactance (through a transformer). At all operational frequencies, the inductance and capacitance of the full SQUID array (both coupled and parasitic) will be small as compared to the inductance of the coaxial pickup \cite{DMRmSquids}. The coupling to the dc SQUID is also small enough that damping (i.e. resistance) from both active and passive loss in the SQUID is subdominant to the loss in the electrons in the anomalous skin depth of the coaxial pickup \cite{DMRmSquids}. 

Following the conventions in \cite{Chaudhuri:2018rqn}, the SQUID noise is described by a parameter $\eta$ as the ratio of the noise temperature to a half-photon of quantum noise:

\begin{equation}
\eta(\nu_0)\equiv\frac{kT_{N}^{min}(\nu_0)}{h\nu_0/2} \geq 1.
\label{eq:eta}
\end{equation}
An amplifier capable of operating at the Standard Quantum Limit (SQL) at optimal noise match would have $\eta=1$, corresponding to a minimum of a half photon of added noise on resonance. Ultimately, the quantity that enters the scan rate is $\bar{\mathcal{G}}[\nu_r,T,\eta(\nu_r)]$, which encodes all the noise physics of the system (described in more detail in \cite{Chaudhuri:2018rqn,Brouwer:2022DMRm}).

\begin{figure}
    \centering
    \includegraphics[width=.44\textwidth]{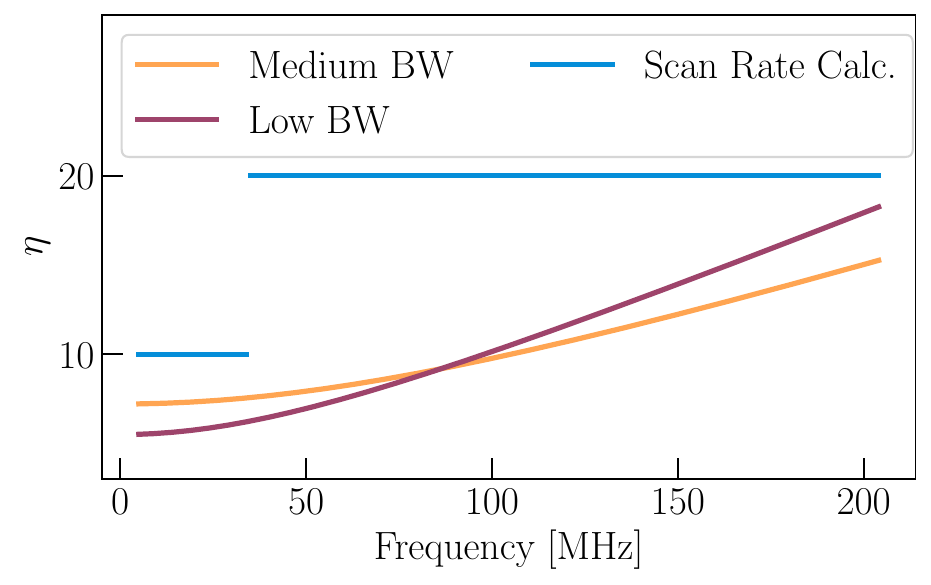}
    \caption{The frequency dependent variation of $\eta$ for the Medium (orange) and Low BW (purple) SQUID chains over the DMRadio-m$^3$ frequency range as presented in in Table \ref{tab:scantime}. The Low BW SQUID chain is optimal in the 5-75 MHz range while the Medium BW SQUID chain is optimal in the 75-200 MHz range. A conservative estimate for these values has been used instead in the scan rate analysis as indicated by the blue line. This blue line is $\eta=10$ for 10-30 MHz and $\eta=20$ for 30-200 MHz. The scan rate shown in Figure \ref{fig:scanrate} utilizes these conservative values. 
    }
    \label{fig:etas}
\end{figure}

Details of the SQUID amplifiers are presented in Table \ref{tab:1stSQUIDparameters} and the value of $\eta(\nu_0)$ for the two different SQUID chains is shown Figure \ref{fig:etas}. These results are described in detail in \cite{DMRmSquids} where an analysis is presented for previously realized first-stage SQUIDs whose behavior is accurately modeled by Tesche-Clarke theory, coupled to prototype second-stage dc SQUID arrays. These results refer the dc SQUID noise to a high-$Q$ resonator and the analysis takes into account bandwidth, SQUID backaction noise, imprecision noise, and their correlations. The analysis in \cite{DMRmSquids} considers a broader set of amplifier chains, including some of which are only optimal outside of the frequency ranges covered by \DMRm{}.



\begin{table}[]
    \centering
\begin{tabular}{c|c|c}
 & Low BW & Medium BW                                        \\ \hline
\begin{tabular}[c]{@{}c@{}} \# of SQUIDs in \\ Second Stage  \end{tabular}   &   144 ($48\times 3)$ &  64 $(32\times 2)$                                                    \\ \hline
 $R_\text{dyn}$     & 50\,$\Omega$  &    50\,$\Omega$                               \\ \hline
Freq. range      & 
5--75\,MHz &    75--200\,MHz                          \\   \hline
$\eta$ range    & 5--8 & 8--15                           \\                                                 
\end{tabular}
    \caption{Parameters of the two different two-stage SQUID amplifier chains used in \DMRm{}: Low and Medium bandwidth (BW). The first stage consists of a single SQUID. The second stage consists of an array of SQUIDs using one common SQUID design. Details and experimentally measured parameters are in \cite{DMRmSquids}. The input coils of all SQUIDs in the second stage are wired in series. The output of the second stage has $N_{\rm ser}$ SQUIDs in series, and $N_{\rm par}$ SQUIDs in parallel, for a total number of $N_{\rm 2}$ SQUIDs, denoted in the table as $N_{\rm 2} (N_{\rm ser} \times N_{\rm par})$. The output dynamic resistance 
    $R_\text{dyn}$ of the second-stage SQUID is listed at the bias point, and is matched to a $50\ \Omega$ coaxial line and room-temperature preamplifier input impedance. The frequency range specifies which of the two SQUID chains is used at each frequency. The last row contains values for $\eta$ (defined in Equation \ref{eq:eta}), the total noise of the two-stage SQUID, including the referred room-temperature preamplifier noise. It incorporates imprecision and backaction noise, and their correlations.  Figure \ref{fig:etas} illustrates $\eta$ for both SQUID chains over the operational frequencies of DMRadio-m$^3$. Frequency range and $\eta$, the last two rows of the table, are the only parameters needed for this analysis.}

\label{tab:1stSQUIDparameters}
\end{table}

\section{Science Reach}
\label{sec:ScienceReach}
\begin{figure*}[ht]
    \centering
    \includegraphics[width=.99\textwidth]{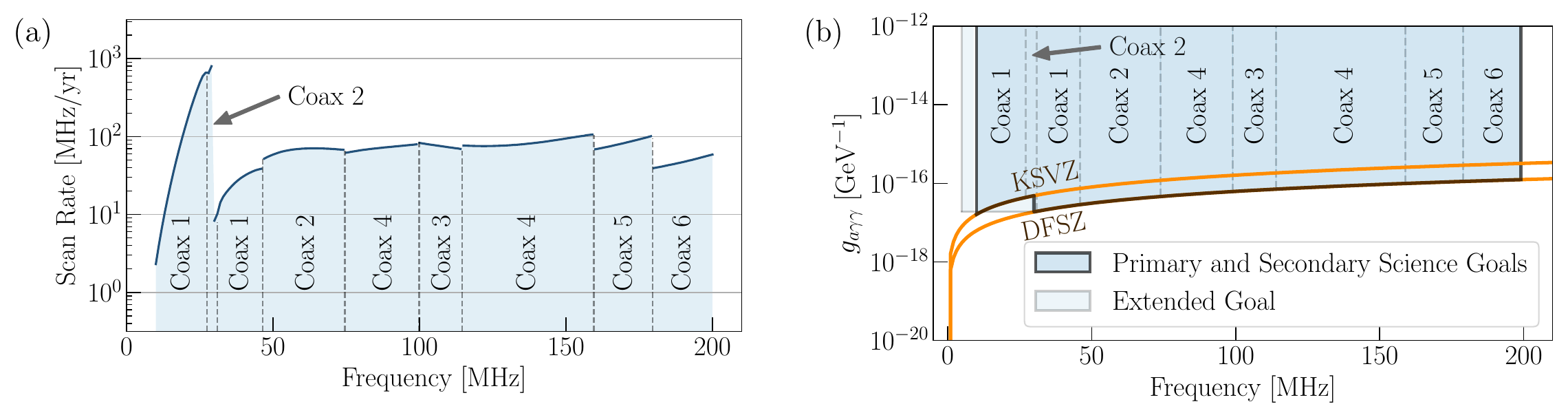}
    \caption{(a) The baseline scan rate for \DMRm{} across the primary and secondary science goals using the set of six coaxial pickups. The primary science goal is to search for axions at DFSZ coupling at 30--200 MHz; the secondary goal is to search for KSVZ axions at 10--30 MHz. Above 75 MHz, the High BW SQUID amplifier is used. 
    A conservative value of $\eta=20$ is also shown in the 30-87 MHz range where the predicted $\eta$ shown in Fig. \ref{fig:etas} is otherwise lower than 20. These scan rates all assume the constant $6\times$ increase in $Q$ due to the cryogenic increase in the conductivity of copper. Scan times are available at \cite{alshirawi_2025_15149989}  (b) The sensitivity of \DMRm{} for the primary and secondary science goals, as well as for the extended  goal which is defined as axions with a coupling of $g_{a\gamma\gamma}=g_{a\gamma\gamma,\text{DFSZ}}(30\text{ MHz})=1.87\times10^{-17}\text{ GeV}^{-1}$ over 5--30 MHz. The two orange lines show the axion-photon coupling of KSVZ (top line) and DFSZ (bottom line) axions.
    The coaxes used for each region are also shown.
    }
    \label{fig:scanrate}
\end{figure*}

The primary science goal of \DMRm{} is to search for DFSZ axions across the 30--200 MHz region. Using the results of the previous sections, we first calculate the scan rate across this frequency range. As indicated in Figure \ref{fig:scanrate}, the full scan uses six different coaxes with their corresponding tuning mechanisms.

\begin{table}[]
    \centering
\begin{tabular}{c|c|c}
 & Sensitivity \& Range &   \begin{tabular}[c]{@{}c@{}} $3\sigma$ Live Scan Time: \\ Baseline (Scaled)  \end{tabular}                                          \\ \hline
\begin{tabular}[c]{@{}c@{}} Primary \\ Science Goal  \end{tabular}   &   DFSZ; 30--200\,MHz    & 2.9  yr (2.6 yr)                                                        \\ \hline
\begin{tabular}[c]{@{}c@{}} Secondary \\ Science Goal  \end{tabular}    & KSVZ; 10--30\,MHz   &  1.0 yr (0.6 yr)                               \\ \hline
\rule{0pt}{4ex}  \begin{tabular}[c]{@{}c@{}} Extended \\  Goal  \end{tabular}       & 
\begin{tabular}[c]{@{}c@{}}
$1.87\times10^{-17}\text{ GeV}^{-1}$;\\  5--30\,MHz\end{tabular}
    &   3.5 yr (2.1 yr)                           \\                                                     
\end{tabular}
    \caption{The calculated experimental $3\sigma$ live times for the primary and secondary science goals as well as the extended goal of \DMRm{}. The extended goal live time is the additional time required to achieve this goal assuming the secondary science goal has already been achieved. For the baseline live scan time, this table assumes SQUID amplifiers with  $\eta=10$ for  5--30\,MHz, $\eta=20$ for 30--200\,MHz and the flat factor of $6\times$  increase in $Q(\nu)$ as discussed in \S \ref{sec:subExtractingQ}. Using the $Q$ enhancement factor described in Appendix \ref{sec:AppRRR}, which takes into account the scaling of $Q(\nu)\sim \nu^{-1/6}$ and magnetoresistance effects, we calculate the more aggressive live times shown in parenthesis as ``Scaled.''
}
    \label{tab:scantime}
\end{table}

For the primary science goal, we calculate a ``3$\sigma$ live time,'' which we define as the live time required so that a DFSZ axion is expected to present a 3$\sigma$ signal  (SNR=3 in Eq. \ref{eq:scanrate}). This live time does not include any time spent rescanning.
The ADMX experiment determines rescan requirements with frequentist inference \cite{ADMX:2020hay}, and spends approximately 30\% of its live time on rescans \cite{ADMXprivatecom}. The HAYSTAC experiment uses a Bayesian analysis framework to plan rescans and determine its final sensitivity \cite{Palken:2020wgs}. A detailed rescan strategy and analysis framework for \DMRm{} will be the subject of a future paper. 


The SQUID amplifier is exchanged once at 75\,MHz. In the 5--75\,MHz range, the 
Low BW SQUID amplifier in Table \ref{tab:1stSQUIDparameters} is used; in the 55--200\,MHz range, the High BW SQUID amplifier is used. However, for the scan times shown in Table \ref{tab:scantime} and scan rates shown in Fig. \ref{fig:scanrate}, a more conservative assumption of $\eta=10$ for the 10-30 MHz range, and $\eta=20$ for the 30-200 MHz range is used. Furthermore, across the entire frequency range we consider the two cases for the enhanced quality factor at cryogenic temperatures as discussed in Appendix \ref{sec:AppRRR}: a conservative estimate of a $6\times$ increase of the RT conductivity of copper across all frequencies and a more ambitious increase due to the frequency dependent effects on the resistivity.

For the primary science goal of \DMRm{}, using the scan rates shown in Figure \ref{fig:scanrate}a, the $3\sigma$ live scan time is 2.9 years.  Using the more ambitious frequency-dependent resistivity for copper across the entire frequency range, the resulting $3\sigma$ live time is 2.6 years. All relevant $3\sigma$ live times are shown in Table \ref{tab:scantime}.

The secondary science goal of \DMRm{} covers axions with KSVZ couplings in the 10--30 MHz range. The total live scan time for such axions is 1.0 years, with the corresponding scan rate also shown in Figure \ref{fig:scanrate}a. The total primary and secondary science goal is presented in Figure \ref{fig:scanrate}b.

We also present an extended  goal that searches for axions in the 5--30 MHz range down to a constant axion-photon coupling of 
$g_{a\gamma\gamma,\text{DFSZ}}(\nu=30\text{ MHz})=1.87\times10^{-17}\text{ GeV}^{-1}$, as shown in \ref{fig:scanrate}b. This represents a more ambitious scan plan which excludes more parameter space but at a longer live scan time. This goal requires an additional baseline $3\sigma$ live scan time of 3.5 years for a $g_{a\gamma\gamma,\text{DFSZ}}(\nu=30\text{ MHz})=1.87\times10^{-17}\text{ GeV}^{-1}$ axion. This $3\sigma$ live time assumes that the secondary science goal parameter space has already been covered  and that that data can be used in the extended  goal scan. In all cases, the time for rescans is not included.

\section{Conclusion}
In this work, we have presented a model of the electromagnetic performance of \DMRm{}. Because the Compton wavelength scanned by this experiment approaches twice the detector size, we have applied a full numerical model of the modified Maxwell's equation response of the \DMRm{} detector, which is valid over the full frequency range. We have extracted the impedance of the structure, modeled it as a different series RLC circuit at each frequency of interest, and have determined the voltage signal $V(\nu)$ induced by the axions. Using the noise properties of the SQUID readout amplifiers \cite{DMRmSquids}, we have determined the live scan times.


\DMRm{} operates over a significant range of well motivated QCD axion parameter space. We have determined (see Table \ref{tab:scantime}) the $3\sigma$ live time for DFSZ axions in the 30--200 MHz range (2.6--2.9 years) and KSVZ axions in the 10-30 MHz range (0.6--1.0 years), with an extended goal to further to search for axions at $g_{a\gamma\gamma}=1.87\times10^{-17}\text{ GeV}^{-1}$ over 5--30 MHz. 


\begin{acknowledgments}
The authors acknowledge support for \DMRm as part of the DOE Dark Matter New Initiatives program under SLAC FWP 100559. Members of the \DMR Collaboration acknowledge support from the NSF under awards 2110720 and 2014215. Stanford University and UC Berkeley gratefully acknowledge support from the Gordon and Betty Moore Foundation, grant number 7941, and additional support from the Heising-Simons Foundation. C.~Bartram acknowledges support from the Panofsky Fellowship at the SLAC National Accelerator Laboratory. S.~Chaudhuri acknowledges support from the R.~H.~Dicke Postdoctoral Fellowship and Dave Wilkinson Fund at Princeton University. J.~W.~Foster was supported by a Pappalardo Fellowship. J.~T.~Fry is supported by the National Science Foundation Graduate Research Fellowship under Grant No. 2141064. P.~W.~Graham acknowledges support from the Simons Investigator Award no. 824870 and the Gordon and Betty Moore Foundation Grant no. 7946. Y.~Kahn was supported in part by DOE grant DE-SC0015655.  B.~R.~Safdi and J.~N.~Benabou were supported  in part  by  the  DOE  Early  Career  Grant  DESC0019225. C.~P.~Salemi is supported by the Kavli Institute for Particle Astrophysics and Cosmology Porat Fellowship.

\end{acknowledgments}


\appendix
\section{Shielded box considerations}
\label{sec:Box}

As shown in Fig. \ref{fig:circmodel}a, the shielded box, which is shaped like a cylindrical pillbox and houses the tuning elements and readout electronics, is located in the magnet's low-field region directly above the straight coaxial section. This box shields the sensitive SQUIDs from the large dc magnetic field and from external interference. Although the size of the low-field region is set by the magnet and bucking coils, the dimensions of this shielded box are free to vary in the optimization of the detector geometry. We show here that variations in the dimensions of this shielded box around the design value have a negligible effect on the pickup's response to an axion signal.

\begin{figure}[ht]
    \centering
    \includegraphics[width=.48\textwidth]{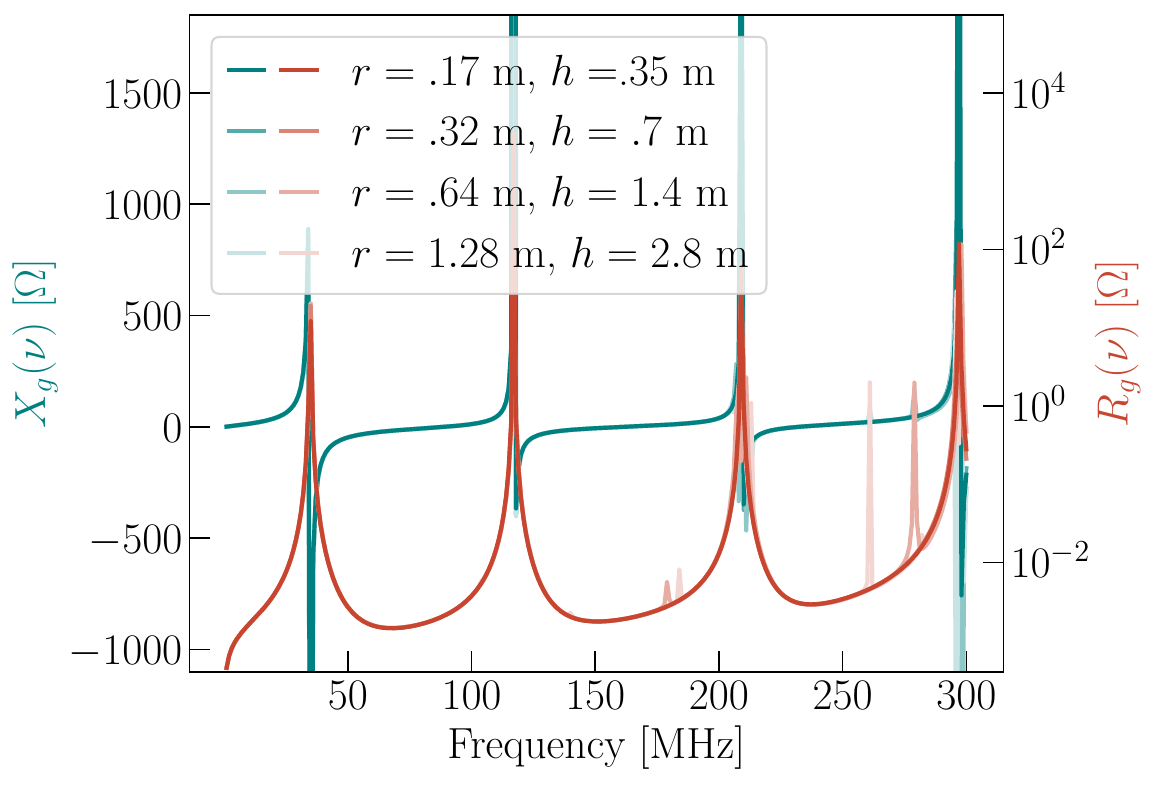}
    \caption{Reactive (teal) and resistive (red) curves for a coaxial pickup of height 1.4 m with shielded box sizes of varying radius $r$ and height $h$. The curves, with the exception of a few lines above 175 MHz, are nearly identical over 300 MHz thus showing that the effects of the box are not significant for \DMRm{}.}
    \label{fig:Boxes}
\end{figure}

Because the shielded box is a right circular cylinder, its interior supports both TE and TM modes. At  large box sizes, these modes begin to intrude into the \DMRm{} science frequencies. For small box sizes, capacitance between the walls of the box and the inner conductor of the coax shorts out signal currents. To study the effect of box sizes on signal response, simulations were run in \texttt{COMSOL} with varying box dimensions. The result of four of these simulations is shown in Fig. \ref{fig:Boxes}. The effect of the box on the coax's impedance is negligible except at a few specific frequencies which represent TE or TM modes in the two largest boxes studied ($r=0.64~\text{m}, h=1.4~\text{m}$ and $r=1.28~\text{m}, h=2.8~\text{m}$). These RF modes are found slightly higher ($\sim1$ MHz) than the frequencies predicted by analytical formulas for right circular cylinders due to coupling to the impedance of the coax. Shielded boxes smaller than $r=0.17~\text{m}\text{ by } h=0.35~\text{m}$ were not studied since they are too small to contain our electronics. The selected dimensions of the shielded box insure that none of these modes intrude into signal frequencies. Furthermore, the shielded box does not add significant capacitive shunting across the coax.

\section{COMSOL simulation techniques }
\label{sec:AppCOMSOL}

In this appendix, we describe the procedure used to characterize the coaxial pickup's response to an applied axion current. With the RF Module in \texttt{COMSOL} 6.0, we make a 2D axisymmetric model of the detector geometry and define a coaxial lumped port with cable terminal across the open end of the model (points A and B in Fig. \ref{fig:circmodel}). The width of the lumped port is much smaller than the characteristic wavelength of the highest frequency being simulated ($f=200~\text{MHz}$, corresponding to $\lambda\approx150~\text{cm}$), allowing for voltage, $V(\omega)$ and current, $I(\omega)$, to be well-defined across the port for all frequencies. The characteristic impedance of this lumped port in the simulation is set to $Z_{\text{ref}} = 50\times 10^9~\Omega$ to ensure that the port impedance has a minimal effect on the measured impedance of the coaxial pickup. This value is chosen such that the port acts as a current source. An alternative simulation using a circuit terminal for this lumped port eliminates the need to define a characteristic port impedance and gives identical results.
 
 Making use of the Impedance Boundary Condition feature of \texttt{COMSOL}, we define the boundaries of the coaxial pickup to have the surface impedance of metallic copper at room temperature. This feature allows us to calculate loss (and therefore the real part of $Z_p(\omega$)) without solving the electromagnetic equations within a skin depth inside the material, therefore reducing computation time. As described elsewhere, the modeled loss is later adjusted to account for higher conductivity at low temperatures, and the effect of magnetoresistance.
 
 The impedance of the coaxial pickup, $Z_p(\omega)$, can then be identified by exciting an ac voltage across the lumped port in a frequency domain study. $Z_p(\omega)$ is read out as the impedance across the lumped port, referred to as \texttt{emw.Zport\_1} on \texttt{COMSOL}.
 
 The voltage induced by the axion signal across the open end of the coaxial pickup, $V(\omega)$, is found by turning off the wave excitation at the lumped port and introducing an external current density excitation throughout the entire volume of the coaxial pickup. The external current density in \texttt{COMSOL} provides a phase-coherent electric current,  representing the excitation due to an axion. We define an external current density by:
 \begin{equation}
      \mathbf{J_e}(t,\rho,\phi, z)=1\text{ A/m}^2\left(\frac{\mathbf{B}(\rho,\phi,z)}{1\text{ T}}\right)e^{i\omega t},
 \label{eq:extcurrentdens}
 \end{equation}
 where $\mathbf{B}$ is the vector field of the magnetic field, which adjusts the effective axion current density for the strength of the dc magnetic field. We then apply an additional scaling factor that depends on the coupling strength and  dark matter model.
 
 Due to azimuthal symmetry, this field only has components along the $\hat{\rho}$ and $\hat{z}$ directions. $B_\rho$ and $B_z$ are found numerically by modeling the coils of the solenoidal magnet in a separate \texttt{COMSOL} simulation, using the Magnetic Fields interface within the ac/dc Module. Importing the resulting $r$ and $z$ components of the applied magnetic field, which are defined on a grid, and then defining an interpolation for each of the two components, we define $\mathbf{J_e}$ in terms of $B_\rho$ and $B_z$ as above in Eq. \ref{eq:extcurrentdens}.

To extract the axion-induced voltage we use a frequency-domain simulation with the external current density acting as the ac source. The voltage is extracted by integrating the radial electric field along a line across the gap (points A and B in Figure \ref{fig:circmodel}):
\begin{equation}
    \label{eqn:Veqn}V_p(\omega)=\int_A^B E_r dr,
\end{equation}
where $r$ represents the radial component in the 2D axisymmetric situation. The integration path is always much shorter than a wavelength, which ensures that Eq. \ref{eqn:Veqn} is valid.
 

\section{Copper coax surface resistance}
\label{sec:AppRRR}
The statistical noise limit of the experiment improves with the quality factor, so it is
important to accurately calculate loss and thus resonator $Q$. 

The surface resistance of copper at room temperature is given by 
\begin{equation}
    Re\left\{ Z_{\rm RT}\right\}=\sqrt{\frac{\omega \mu_0}{2 \sigma_{\rm RT}}},
    \label{eq:NormSkinEffect}
\end{equation}
where $\sigma_{\rm RT}$ is the room-temperature conductance \cite{jackson_classical_1999}.

In our \texttt{COMSOL} models, we assume standard room-temperature, high-conductivity copper, and then increase the modeled $Q$ by a factor $Q_{\rm rat}(\omega)$, which is the ratio of surface resistance of copper at room temperature to its cryogenic value at angular frequency $\omega$.

At temperatures well below 4~K, the dc resistivity decreases by a large factor. A dc Residual Resistivity Ratio (RRR) of above 400 can be achieved for copper. However, at higher frequencies and $T<4$~K, the resonator loss decreases by a much smaller factor due to the anomalous skin-effect \cite{PIPPARD19541}. In this extreme anomalous limit, the surface resistance of the conductor is given by
\begin{equation}
    Re\left\{ Z_{\rm cold}\right\}=\frac{8}{9}
    \left(\frac{\sqrt{3} \lambda_\text{mfp} \omega^2 \mu_0^2} 
{16 \pi \sigma_{\rm cold}}\right)^{1/3},
    \label{eq:SurfImp}
\end{equation}
where $\sigma_{\rm cold}$ is dc conductivity at low temperatures, and $\lambda_{\text{mfp}}$ is the mean free path of the electrons.

The ratio of room-temperature surface resistance (Eq. \ref{eq:NormSkinEffect}) to extreme anomalous low-temperature surface resistance (Eq. \ref{eq:SurfImp}), and thus the ratio of the achieved $Q$ between room and cryogenic temperatures, is
\begin{equation}
    Q_{\rm rat}(\omega) \equiv \frac{Re\left\{ Z_{\rm RT}\right\}}{Re\left\{ Z_{\rm cold}\right\}}
    \propto \nu^{-1/6}.
    \label{eq:Qscale}
\end{equation}
The scaling of the ratio of loss $Q_{\rm rat}$ with frequency is found to hold true over a large range of frequency, with an example achieved value of $Q_{\rm rat}=4.8$ at 2.85\,GHz \cite{Cahill:2016uui}. This study of resonator $Q$ is consistent with extensive experimental measurements conducted in cryogenic copper cavities \cite{peng2000cryogenic,Backes2021} and analysis of cryogenic copper RF conductance \cite{finger2008microwave}.
Using the $\nu^{-1/6}$ scaling with the reference value of $Q_{\rm rat}=4.8$ at 2.85\,GHz, the predicted value of $Q_{\rm rat}$ is  13.8 at 5\,MHz, 10.3 at 30\,MHz, and 7.5 at 200\,MHz. The HAYSTAC experiment has achieved a $Q_\text{rat}$ of approximately 4 at 1\,GHz \cite{simanovskaia2021} for its copper cavity and the CAPP experiment has demonstrated a $Q_\text{rat}$ between 2 and 3 for frequencies of 4.7-5.0\,GHz \cite{PhysRevD.106.092007} for its copper cavity. Using the HAYSTAC and CAPP numbers as the reference values for the $\nu^{-1/6}$ scaling yields $Q_\text{rat}$  values between 5 and 7 at 30 MHz.

Magnetoresistance can also change the value of $Q$ achieved in \DMRm{}. The magnetoresistance is dominated by the longitudinal (axial) component of the dc magnetic field in the copper coax endcaps, which is orthogonal to the direction of current flow, and to a lesser extent by the transverse (radial) component in the cylindrical walls.  In our case, the average transverse field contributing  to magnetoresistance is $\sim1$\,T. The low-temperature dc magnetoresistance of OFHC copper with $\text{RRR}=100$ at 1\,T  is $\Delta\rho/\rho\approx10$--30\% \cite{benz1969magnetoresistance,fickett1972magnetoresistivity}, consistent with Kohler’s rule \cite{kohler1938magnetischen}. For the ac magnetoresistance, in the \DMRm{}  frequency range at $B=1$\,T, the cyclotron radius is of the same order as the skin depth and mean free path, and the ac magnetoresistance is expected to be significantly lower than the dc value of $\sim10$--30\%  \cite{rogers1988anomalous,ahn2017magnetoresistance}.  It should be noted that the HAYSTAC and CAPP experiments use stronger magnetic fields, and should have somewhat different magnetoresistance.

\begin{figure}[ht]
    \centering
    \includegraphics[width=.48\textwidth]{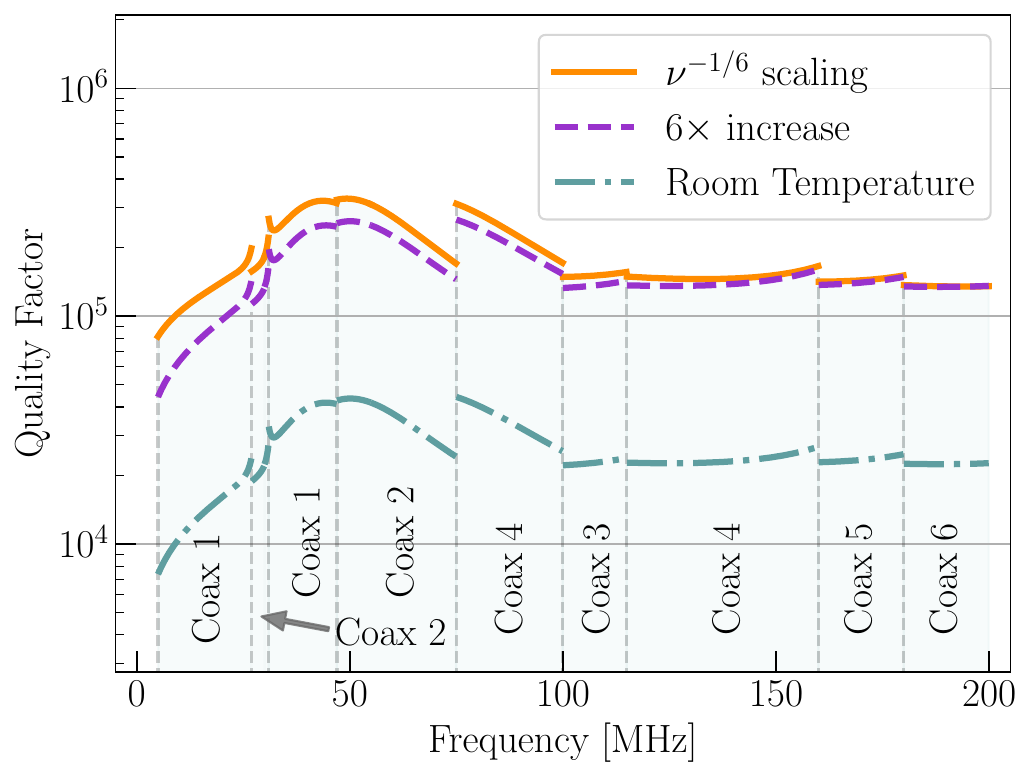}
    \caption{The total quality factor $Q(\nu)$ for the \DMRm{} experiment. This quality factor combines the contributions from the coaxial copper pickups as well as from capacitive or inductive tuning elements. We assume that the loss in this system is dominated by the normal electrons in the copper of the coaxial pickup. Three curves for $Q(\nu)$ are shown. In dot-dashed green we show the quality factor using the room temperature conductivity of copper. In dashed purple, we show the conservative $6\times$ increase in the quality factor at cryogenic temperatures. In orange we show the frequency dependent $\nu^{-1/6}$ scaling with the effects of magnetoresistance.}
    \label{fig:Qfactor}
\end{figure}

In the main text, we extract the main results assuming a conservative baseline constant value of $Q_\text{rat}=6$ across the frequency range. However, in Table \ref{tab:scantime} we include the results assuming the scaling in Eq. \ref{eq:Qscale} with a reference value of 4.8 at 2.85\,GHz \cite{Cahill:2016uui} as well as a 20\% decrease in $Q$ due to the effects of magnetoresistance. These combine to adjust the $Q$ by a factor of 11.1 at 5\,MHz, 9.9 at 10\,MHz, 8.2 at 30\,MHz, and 6.0
 at 200\,MHz which, in turn, lead to a decrease in live scan time (shown as the scaled values in Table \ref{tab:scantime}). These values for $Q(\nu)$ are shown in Figure \ref{fig:Qfactor}.  They include the inductive or capacitive contributions from the tuning elements and are shown for three cases: (a) the room temperature conductivity of copper as modeled by \texttt{COMSOL}, (b) using the constant conservative $6\times$ increase which provides the baseline scan time in Table \ref{tab:scantime}, and (c) for the $\nu^{-1/6}$-dependent scaling which provides the scaled scan time in Table \ref{tab:scantime}. The loss is assumed to be dominated by the normal electrons in the copper.

\section{Bulk Electron Shuttling}
\label{sec:AppShuttling}

In section \ref{sec:equivalent_RLC}, we mention an effect arising from the axion-modified form of Maxwell's equations that leads to current nodes shown schematically in the coax in Figure \ref{fig:circmodel}b. We refer to this phenomenon as \textit{bulk electron shuttling} (BES). In this appendix, we derive this phenomenon in the bulk of a conductor many skin depths thick, with electrical conductivity $\sigma$. This is a relevant limit to many axion experiments, including \DMRm{}, in which the dc magnetic field, and thus the axion signal, penetrates bulk conductors.

The axion-modified form of Amp\`ere's law inside of an electrical conductor is given by:
\begin{equation}
    \nabla\times \mathbf{B}=\mu_0\epsilon_0\frac{\partial \mathbf{E}}{\partial t}+\mu_0 \sigma \mathbf{E}+\mu_0\mathbf{J}_\text{eff},
    \label{eq:AppAmpere}
\end{equation}
where $\mathbf{J}_\text{eff}$ represents the effective axion-induced electrical current presented in Equation \ref{eq:AxCurrentDens}. We have used Ohm's law to replace the electron current density in the conductor:
\begin{equation}
\mathbf{J}_e=\sigma\mathbf{E},
\label{eq:AppOhm}
\end{equation}
and we assume that the electrical conductivity is uniform. By taking the curl of Equation \ref{eq:AppAmpere} and using the Maxwell-Faraday equation:
\begin{equation}
    \nabla\times\mathbf{E}=-\frac{\partial\mathbf{B}}{\partial t},
\end{equation}
we derive the axion-modified wave equation for the electric field:
\begin{equation}        
    \nabla^2\mathbf{E}=\mu_0\epsilon_0\frac{\partial^2\mathbf{E}}{\partial t^2}+\mu_0\sigma\frac{\partial\mathbf{E}}{\partial t}+\mu_0 \frac{\partial\mathbf{J}_\text{eff}}{\partial t}.
    \label{eq:AppC5}
\end{equation}
Examining this equation in the Fourier domain and converting to natural units where $c=\mu_0=1$, we establish a dispersion relation:
\begin{equation}
    \left(-k^2+\omega^2-i\omega\sigma\right)\mathbf{\tilde{E}} =i\omega \tilde{\mathbf{J}}_\text{eff},
    \label{eq:AppDisp}
\end{equation}
where $\tilde{\mathbf{E}}$ and $\tilde{\mathbf{J}}_\text{eff}$ are the Fourier transforms of $\mathbf{E}$ and $\mathbf{J}_\text{eff}$ respectively. We consider which terms are dominant in the operating conditions of \DMRm{}.
In a good electrical conductor such as copper,  a conductivity of $\mathcal{O}(10^{7}\text{ S/m})$
corresponds to a frequency scale in natural units of $\mathcal{O}(10^{18}\text{ Hz})$ and a characteristic wavelength of $\mathcal{O}(100\text{ pm})$. For electric field oscillation frequencies less than $10^{18}$ Hz and spatial variations in the field  larger than hundreds of pm, the left hand side of Eq. \ref{eq:AppDisp} is dominated by the $i\omega\sigma$ term thus resembling a heavily over-damped harmonic oscillator. In this limit, \ref{eq:AppDisp} reduces to: 
\begin{equation}
    \tilde{\mathbf{E}}=-\frac{1}{\sigma}\tilde{\mathbf{J}}_\text{eff}.
\end{equation}
Using Ohm's law (Eq. \ref{eq:AppOhm}), we identify the BES current density, which is equal and opposite to the axion-induced effective current density:
\begin{equation}
    \tilde{\mathbf{J}}_{e} = - \tilde{\mathbf{J}}_\text{eff}.
    \label{Eq:AppBES}
\end{equation}

This formula holds only in the regimes where $\sigma\gg\omega$ and also when $\sigma$ is much larger than the inverse of the length scales associated with the variation in the fields ($\sigma\omega \gg k^2$).  While this derivation has assumed that $\sigma$ is uniform across the bulk, the results derived hold for a spatially varying $\sigma(\mathbf{x})$, as long as the variation occurs on length scales larger than the skin depth of the metal (tens of $\mu$m for copper at the frequencies of interest); this is evident considering that in natural units the skin depth $\delta$ is given by $\delta=\sqrt{2/(\sigma\omega)}$ and hence the corresponding wavenumber $k$ is $k\sim \sqrt{\sigma\omega}$. It is also worth noting that Ohm's Law breaks down  when the mean-free-path of the electrons becomes larger than the wavelength of the signal; for a metal like copper this occurs around $10^{15}$ Hz. Hence, Eq. \ref{Eq:AppBES} only holds for low frequencies and fields with sufficiently high spatial uniformity (varying over length scales large as compared to the skin depth).

The axion effective current generates an ac BES current of equal magnitude and opposite direction in the bulk of the copper pickup of \DMRm{}, moving charges across the bulk of the metal, even at depths much larger than the skin depth. This phenomenon is required to understand the current nodes in Figure \ref{fig:circmodel}, since some screening current is ``shuttled'' through the bulk of the copper, in accordance with Eq. \ref{Eq:AppBES}. It is also interesting to note that although these ac currents are moved across the bulk, electromagnetic interference is still shielded by the skin depth of the copper. This is because the ac BES current passing through the bulk deeper than the skin depth, according to Eq.  \ref{Eq:AppBES}, is dependent only on the effective axion current, and is independent of any external electromagnetic excitation (it is a stiff current transfer). This analysis can be extended to BES in a superconductor that is penetrated by dc magnetic flux.


\bibliographystyle{apsrev4-2}
\bibliography{dmradio_bibliography}

\begin{thebibliography}{78}%
\makeatletter
\providecommand \@ifxundefined [1]{%
 \@ifx{#1\undefined}
}%
\providecommand \@ifnum [1]{%
 \ifnum #1\expandafter \@firstoftwo
 \else \expandafter \@secondoftwo
 \fi
}%
\providecommand \@ifx [1]{%
 \ifx #1\expandafter \@firstoftwo
 \else \expandafter \@secondoftwo
 \fi
}%
\providecommand \natexlab [1]{#1}%
\providecommand \enquote  [1]{``#1''}%
\providecommand \bibnamefont  [1]{#1}%
\providecommand \bibfnamefont [1]{#1}%
\providecommand \citenamefont [1]{#1}%
\providecommand \href@noop [0]{\@secondoftwo}%
\providecommand \href [0]{\begingroup \@sanitize@url \@href}%
\providecommand \@href[1]{\@@startlink{#1}\@@href}%
\providecommand \@@href[1]{\endgroup#1\@@endlink}%
\providecommand \@sanitize@url [0]{\catcode `\\12\catcode `\$12\catcode `\&12\catcode `\#12\catcode `\^12\catcode `\_12\catcode `\%12\relax}%
\providecommand \@@startlink[1]{}%
\providecommand \@@endlink[0]{}%
\providecommand \url  [0]{\begingroup\@sanitize@url \@url }%
\providecommand \@url [1]{\endgroup\@href {#1}{\urlprefix }}%
\providecommand \urlprefix  [0]{URL }%
\providecommand \Eprint [0]{\href }%
\providecommand \doibase [0]{https://doi.org/}%
\providecommand \selectlanguage [0]{\@gobble}%
\providecommand \bibinfo  [0]{\@secondoftwo}%
\providecommand \bibfield  [0]{\@secondoftwo}%
\providecommand \translation [1]{[#1]}%
\providecommand \BibitemOpen [0]{}%
\providecommand \bibitemStop [0]{}%
\providecommand \bibitemNoStop [0]{.\EOS\space}%
\providecommand \EOS [0]{\spacefactor3000\relax}%
\providecommand \BibitemShut  [1]{\csname bibitem#1\endcsname}%
\let\auto@bib@innerbib\@empty
\bibitem [{\citenamefont {Peccei}\ and\ \citenamefont {Quinn}(1977{\natexlab{a}})}]{Peccei:1977hh}%
  \BibitemOpen
  \bibfield  {author} {\bibinfo {author} {\bibfnamefont {R.~D.}\ \bibnamefont {Peccei}}\ and\ \bibinfo {author} {\bibfnamefont {H.~R.}\ \bibnamefont {Quinn}},\ }\href {https://doi.org/10.1103/PhysRevLett.38.1440} {\bibfield  {journal} {\bibinfo  {journal} {Phys. Rev. Lett.}\ }\textbf {\bibinfo {volume} {38}},\ \bibinfo {pages} {1440} (\bibinfo {year} {1977}{\natexlab{a}})}\BibitemShut {NoStop}%
\bibitem [{\citenamefont {Peccei}\ and\ \citenamefont {Quinn}(1977{\natexlab{b}})}]{Peccei:1977ur}%
  \BibitemOpen
  \bibfield  {author} {\bibinfo {author} {\bibfnamefont {R.~D.}\ \bibnamefont {Peccei}}\ and\ \bibinfo {author} {\bibfnamefont {H.~R.}\ \bibnamefont {Quinn}},\ }\href {https://doi.org/10.1103/PhysRevD.16.1791} {\bibfield  {journal} {\bibinfo  {journal} {Phys. Rev.}\ }\textbf {\bibinfo {volume} {D16}},\ \bibinfo {pages} {1791} (\bibinfo {year} {1977}{\natexlab{b}})}\BibitemShut {NoStop}%
\bibitem [{\citenamefont {Weinberg}(1978)}]{Weinberg:1977ma}%
  \BibitemOpen
  \bibfield  {author} {\bibinfo {author} {\bibfnamefont {S.}~\bibnamefont {Weinberg}},\ }\href {https://doi.org/10.1103/PhysRevLett.40.223} {\bibfield  {journal} {\bibinfo  {journal} {Phys. Rev. Lett.}\ }\textbf {\bibinfo {volume} {40}},\ \bibinfo {pages} {223} (\bibinfo {year} {1978})}\BibitemShut {NoStop}%
\bibitem [{\citenamefont {Wilczek}(1978)}]{Wilczek:1977pj}%
  \BibitemOpen
  \bibfield  {author} {\bibinfo {author} {\bibfnamefont {F.}~\bibnamefont {Wilczek}},\ }\href {https://doi.org/10.1103/PhysRevLett.40.279} {\bibfield  {journal} {\bibinfo  {journal} {Phys.Rev.Lett.}\ }\textbf {\bibinfo {volume} {40}},\ \bibinfo {pages} {279} (\bibinfo {year} {1978})}\BibitemShut {NoStop}%
\bibitem [{\citenamefont {Abbott}\ and\ \citenamefont {Sikivie}(1983)}]{Abbott:1982af}%
  \BibitemOpen
  \bibfield  {author} {\bibinfo {author} {\bibfnamefont {L.~F.}\ \bibnamefont {Abbott}}\ and\ \bibinfo {author} {\bibfnamefont {P.}~\bibnamefont {Sikivie}},\ }\href {https://doi.org/10.1016/0370-2693(83)90638-X} {\bibfield  {journal} {\bibinfo  {journal} {Phys.~Lett.~B}\ }\textbf {\bibinfo {volume} {120}},\ \bibinfo {pages} {133} (\bibinfo {year} {1983})}\BibitemShut {NoStop}%
\bibitem [{\citenamefont {Preskill}\ \emph {et~al.}(1983)\citenamefont {Preskill} \emph {et~al.}}]{Preskill1983}%
  \BibitemOpen
  \bibfield  {author} {\bibinfo {author} {\bibfnamefont {J.}~\bibnamefont {Preskill}} \emph {et~al.},\ }\href {https://doi.org/http://dx.doi.org/10.1016/0370-2693(83)90637-8} {\bibfield  {journal} {\bibinfo  {journal} {Phys.~Lett.~B}\ }\textbf {\bibinfo {volume} {120}},\ \bibinfo {pages} {127 } (\bibinfo {year} {1983})}\BibitemShut {NoStop}%
\bibitem [{\citenamefont {Dine}\ and\ \citenamefont {Fischler}(1983)}]{Dine1983}%
  \BibitemOpen
  \bibfield  {author} {\bibinfo {author} {\bibfnamefont {M.}~\bibnamefont {Dine}}\ and\ \bibinfo {author} {\bibfnamefont {W.}~\bibnamefont {Fischler}},\ }\href {https://doi.org/http://dx.doi.org/10.1016/0370-2693(83)90639-1} {\bibfield  {journal} {\bibinfo  {journal} {Phys.~Lett.~B}\ }\textbf {\bibinfo {volume} {120}},\ \bibinfo {pages} {137 } (\bibinfo {year} {1983})}\BibitemShut {NoStop}%
\bibitem [{\citenamefont {Borsanyi}\ \emph {et~al.}(2016)\citenamefont {Borsanyi} \emph {et~al.}}]{Borsanyi:2016ksw}%
  \BibitemOpen
  \bibfield  {author} {\bibinfo {author} {\bibfnamefont {S.}~\bibnamefont {Borsanyi}} \emph {et~al.},\ }\href {https://doi.org/10.1038/nature20115} {\bibfield  {journal} {\bibinfo  {journal} {Nature}\ }\textbf {\bibinfo {volume} {539}},\ \bibinfo {pages} {69} (\bibinfo {year} {2016})},\ \Eprint {https://arxiv.org/abs/1606.07494} {arXiv:1606.07494 [hep-lat]} \BibitemShut {NoStop}%
\bibitem [{\citenamefont {Tegmark}\ \emph {et~al.}(2006{\natexlab{a}})\citenamefont {Tegmark} \emph {et~al.}}]{Tegmark:2005dy}%
  \BibitemOpen
  \bibfield  {author} {\bibinfo {author} {\bibfnamefont {M.}~\bibnamefont {Tegmark}} \emph {et~al.},\ }\href {https://doi.org/10.1103/PhysRevD.73.023505} {\bibfield  {journal} {\bibinfo  {journal} {Phys. Rev. D}\ }\textbf {\bibinfo {volume} {73}},\ \bibinfo {pages} {023505} (\bibinfo {year} {2006}{\natexlab{a}})},\ \Eprint {https://arxiv.org/abs/astro-ph/0511774} {arXiv:astro-ph/0511774} \BibitemShut {NoStop}%
\bibitem [{\citenamefont {Hertzberg}\ \emph {et~al.}(2008)\citenamefont {Hertzberg} \emph {et~al.}}]{Hertzberg:2008wr}%
  \BibitemOpen
  \bibfield  {author} {\bibinfo {author} {\bibfnamefont {M.~P.}\ \bibnamefont {Hertzberg}} \emph {et~al.},\ }\href {https://doi.org/10.1103/PhysRevD.78.083507} {\bibfield  {journal} {\bibinfo  {journal} {Phys. Rev. D}\ }\textbf {\bibinfo {volume} {78}},\ \bibinfo {pages} {083507} (\bibinfo {year} {2008})},\ \Eprint {https://arxiv.org/abs/0807.1726} {arXiv:0807.1726 [astro-ph]} \BibitemShut {NoStop}%
\bibitem [{\citenamefont {Co}\ \emph {et~al.}(2016)\citenamefont {Co} \emph {et~al.}}]{Co:2016xti}%
  \BibitemOpen
  \bibfield  {author} {\bibinfo {author} {\bibfnamefont {R.~T.}\ \bibnamefont {Co}} \emph {et~al.},\ }\href {https://doi.org/10.1103/PhysRevD.94.075001} {\bibfield  {journal} {\bibinfo  {journal} {Phys. Rev. D}\ }\textbf {\bibinfo {volume} {94}},\ \bibinfo {pages} {075001} (\bibinfo {year} {2016})},\ \Eprint {https://arxiv.org/abs/1603.04439} {arXiv:1603.04439 [hep-ph]} \BibitemShut {NoStop}%
\bibitem [{\citenamefont {Graham}\ and\ \citenamefont {Scherlis}(2018)}]{Graham:2018jyp}%
  \BibitemOpen
  \bibfield  {author} {\bibinfo {author} {\bibfnamefont {P.~W.}\ \bibnamefont {Graham}}\ and\ \bibinfo {author} {\bibfnamefont {A.}~\bibnamefont {Scherlis}},\ }\href {https://doi.org/10.1103/PhysRevD.98.035017} {\bibfield  {journal} {\bibinfo  {journal} {Phys. Rev. D}\ }\textbf {\bibinfo {volume} {98}},\ \bibinfo {pages} {035017} (\bibinfo {year} {2018})},\ \Eprint {https://arxiv.org/abs/1805.07362} {arXiv:1805.07362 [hep-ph]} \BibitemShut {NoStop}%
\bibitem [{\citenamefont {Takahashi}\ \emph {et~al.}(2018)\citenamefont {Takahashi} \emph {et~al.}}]{takahashi2018qcd}%
  \BibitemOpen
  \bibfield  {author} {\bibinfo {author} {\bibfnamefont {F.}~\bibnamefont {Takahashi}} \emph {et~al.},\ }\href {https://doi.org/10.1103/PhysRevD.98.015042} {\bibfield  {journal} {\bibinfo  {journal} {Phys. Rev. D}\ }\textbf {\bibinfo {volume} {98}},\ \bibinfo {pages} {015042} (\bibinfo {year} {2018})},\ \Eprint {https://arxiv.org/abs/1805.08763} {arXiv:1805.08763 [hep-ph]} \BibitemShut {NoStop}%
\bibitem [{\citenamefont {Di~Luzio}\ \emph {et~al.}(2020)\citenamefont {Di~Luzio} \emph {et~al.}}]{DiLuzio:2020wdo}%
  \BibitemOpen
  \bibfield  {author} {\bibinfo {author} {\bibfnamefont {L.}~\bibnamefont {Di~Luzio}} \emph {et~al.},\ }\href {https://doi.org/10.1016/j.physrep.2020.06.002} {\bibfield  {journal} {\bibinfo  {journal} {Phys. Rept.}\ }\textbf {\bibinfo {volume} {870}},\ \bibinfo {pages} {1} (\bibinfo {year} {2020})},\ \Eprint {https://arxiv.org/abs/2003.01100} {arXiv:2003.01100 [hep-ph]} \BibitemShut {NoStop}%
\bibitem [{\citenamefont {Wise}\ \emph {et~al.}(1981)\citenamefont {Wise} \emph {et~al.}}]{Wise:1981ry}%
  \BibitemOpen
  \bibfield  {author} {\bibinfo {author} {\bibfnamefont {M.~B.}\ \bibnamefont {Wise}} \emph {et~al.},\ }\href {https://doi.org/10.1103/PhysRevLett.47.402} {\bibfield  {journal} {\bibinfo  {journal} {Phys. Rev. Lett.}\ }\textbf {\bibinfo {volume} {47}},\ \bibinfo {pages} {402} (\bibinfo {year} {1981})}\BibitemShut {NoStop}%
\bibitem [{\citenamefont {Ballesteros}\ \emph {et~al.}(2017)\citenamefont {Ballesteros} \emph {et~al.}}]{Ballesteros:2016xej}%
  \BibitemOpen
  \bibfield  {author} {\bibinfo {author} {\bibfnamefont {G.}~\bibnamefont {Ballesteros}} \emph {et~al.},\ }\href {https://doi.org/10.1088/1475-7516/2017/08/001} {\bibfield  {journal} {\bibinfo  {journal} {JCAP}\ }\textbf {\bibinfo {volume} {08}},\ \bibinfo {pages} {001}},\ \Eprint {https://arxiv.org/abs/1610.01639} {arXiv:1610.01639 [hep-ph]} \BibitemShut {NoStop}%
\bibitem [{\citenamefont {Ernst}\ \emph {et~al.}(2018)\citenamefont {Ernst} \emph {et~al.}}]{Ernst:2018bib}%
  \BibitemOpen
  \bibfield  {author} {\bibinfo {author} {\bibfnamefont {A.}~\bibnamefont {Ernst}} \emph {et~al.},\ }\href {https://doi.org/10.1007/JHEP02(2018)103} {\bibfield  {journal} {\bibinfo  {journal} {JHEP}\ }\textbf {\bibinfo {volume} {02}},\ \bibinfo {pages} {103}},\ \Eprint {https://arxiv.org/abs/1801.04906} {arXiv:1801.04906 [hep-ph]} \BibitemShut {NoStop}%
\bibitem [{\citenamefont {Di~Luzio}\ \emph {et~al.}(2018)\citenamefont {Di~Luzio} \emph {et~al.}}]{DiLuzio:2018gqe}%
  \BibitemOpen
  \bibfield  {author} {\bibinfo {author} {\bibfnamefont {L.}~\bibnamefont {Di~Luzio}} \emph {et~al.},\ }\href {https://doi.org/10.1103/PhysRevD.98.095011} {\bibfield  {journal} {\bibinfo  {journal} {Phys. Rev. D}\ }\textbf {\bibinfo {volume} {98}},\ \bibinfo {pages} {095011} (\bibinfo {year} {2018})},\ \Eprint {https://arxiv.org/abs/1807.09769} {arXiv:1807.09769 [hep-ph]} \BibitemShut {NoStop}%
\bibitem [{\citenamefont {Ernst}\ \emph {et~al.}(2019)\citenamefont {Ernst} \emph {et~al.}}]{Ernst:2018rod}%
  \BibitemOpen
  \bibfield  {author} {\bibinfo {author} {\bibfnamefont {A.}~\bibnamefont {Ernst}} \emph {et~al.},\ }\href {https://doi.org/10.22323/1.347.0054} {\bibfield  {journal} {\bibinfo  {journal} {PoS}\ }\textbf {\bibinfo {volume} {CORFU2018}},\ \bibinfo {pages} {054} (\bibinfo {year} {2019})},\ \Eprint {https://arxiv.org/abs/1811.11860} {arXiv:1811.11860 [hep-ph]} \BibitemShut {NoStop}%
\bibitem [{\citenamefont {Fileviez~P\'erez}\ \emph {et~al.}(2019)\citenamefont {Fileviez~P\'erez} \emph {et~al.}}]{FileviezPerez:2019fku}%
  \BibitemOpen
  \bibfield  {author} {\bibinfo {author} {\bibfnamefont {P.}~\bibnamefont {Fileviez~P\'erez}} \emph {et~al.},\ }\href {https://doi.org/10.1007/JHEP11(2019)093} {\bibfield  {journal} {\bibinfo  {journal} {JHEP}\ }\textbf {\bibinfo {volume} {11}},\ \bibinfo {pages} {093}},\ \Eprint {https://arxiv.org/abs/1908.01772} {arXiv:1908.01772 [hep-ph]} \BibitemShut {NoStop}%
\bibitem [{\citenamefont {Fileviez~P\'erez}\ \emph {et~al.}(2020)\citenamefont {Fileviez~P\'erez} \emph {et~al.}}]{FileviezPerez:2019ssf}%
  \BibitemOpen
  \bibfield  {author} {\bibinfo {author} {\bibfnamefont {P.}~\bibnamefont {Fileviez~P\'erez}} \emph {et~al.},\ }\href {https://doi.org/10.1007/JHEP01(2020)091} {\bibfield  {journal} {\bibinfo  {journal} {JHEP}\ }\textbf {\bibinfo {volume} {01}},\ \bibinfo {pages} {091}},\ \Eprint {https://arxiv.org/abs/1911.05738} {arXiv:1911.05738 [hep-ph]} \BibitemShut {NoStop}%
\bibitem [{\citenamefont {Svrcek}\ and\ \citenamefont {Witten}(2006)}]{Svrcek:2006yi}%
  \BibitemOpen
  \bibfield  {author} {\bibinfo {author} {\bibfnamefont {P.}~\bibnamefont {Svrcek}}\ and\ \bibinfo {author} {\bibfnamefont {E.}~\bibnamefont {Witten}},\ }\href {https://doi.org/10.1088/1126-6708/2006/06/051} {\bibfield  {journal} {\bibinfo  {journal} {JHEP}\ }\textbf {\bibinfo {volume} {06}},\ \bibinfo {pages} {051}},\ \Eprint {https://arxiv.org/abs/hep-th/0605206} {arXiv:hep-th/0605206} \BibitemShut {NoStop}%
\bibitem [{\citenamefont {Green}\ and\ \citenamefont {Schwarz}(1984)}]{Green:1984sg}%
  \BibitemOpen
  \bibfield  {author} {\bibinfo {author} {\bibfnamefont {M.~B.}\ \bibnamefont {Green}}\ and\ \bibinfo {author} {\bibfnamefont {J.~H.}\ \bibnamefont {Schwarz}},\ }\href {https://doi.org/10.1016/0370-2693(84)91565-X} {\bibfield  {journal} {\bibinfo  {journal} {Phys. Lett. B}\ }\textbf {\bibinfo {volume} {149}},\ \bibinfo {pages} {117} (\bibinfo {year} {1984})}\BibitemShut {NoStop}%
\bibitem [{\citenamefont {Conlon}(2006)}]{Conlon:2006tq}%
  \BibitemOpen
  \bibfield  {author} {\bibinfo {author} {\bibfnamefont {J.~P.}\ \bibnamefont {Conlon}},\ }\href {https://doi.org/10.1088/1126-6708/2006/05/078} {\bibfield  {journal} {\bibinfo  {journal} {JHEP}\ }\textbf {\bibinfo {volume} {05}},\ \bibinfo {pages} {078}},\ \Eprint {https://arxiv.org/abs/hep-th/0602233} {arXiv:hep-th/0602233} \BibitemShut {NoStop}%
\bibitem [{\citenamefont {Acharya}\ \emph {et~al.}(2010)\citenamefont {Acharya} \emph {et~al.}}]{Acharya:2010zx}%
  \BibitemOpen
  \bibfield  {author} {\bibinfo {author} {\bibfnamefont {B.}~\bibnamefont {Acharya}} \emph {et~al.},\ }\href {https://doi.org/10.1007/JHEP11(2010)105} {\bibfield  {journal} {\bibinfo  {journal} {JHEP}\ }\textbf {\bibinfo {volume} {11}},\ \bibinfo {pages} {105}},\ \Eprint {https://arxiv.org/abs/1004.5138} {arXiv:1004.5138 [hep-th]} \BibitemShut {NoStop}%
\bibitem [{\citenamefont {Ringwald}(2014)}]{Ringwald:2012cu}%
  \BibitemOpen
  \bibfield  {author} {\bibinfo {author} {\bibfnamefont {A.}~\bibnamefont {Ringwald}},\ }\href {https://doi.org/10.1088/1742-6596/485/1/012013} {\bibfield  {journal} {\bibinfo  {journal} {J. Phys. Conf. Ser.}\ }\textbf {\bibinfo {volume} {485}},\ \bibinfo {pages} {012013} (\bibinfo {year} {2014})},\ \Eprint {https://arxiv.org/abs/1209.2299} {arXiv:1209.2299 [hep-ph]} \BibitemShut {NoStop}%
\bibitem [{\citenamefont {Cicoli}\ \emph {et~al.}(2012)\citenamefont {Cicoli}, \citenamefont {Goodsell},\ and\ \citenamefont {Ringwald}}]{Cicoli:2012sz}%
  \BibitemOpen
  \bibfield  {author} {\bibinfo {author} {\bibfnamefont {M.}~\bibnamefont {Cicoli}}, \bibinfo {author} {\bibfnamefont {M.}~\bibnamefont {Goodsell}},\ and\ \bibinfo {author} {\bibfnamefont {A.}~\bibnamefont {Ringwald}},\ }\href {https://doi.org/10.1007/JHEP10(2012)146} {\bibfield  {journal} {\bibinfo  {journal} {JHEP}\ }\textbf {\bibinfo {volume} {10}},\ \bibinfo {pages} {146}},\ \Eprint {https://arxiv.org/abs/1206.0819} {arXiv:1206.0819 [hep-th]} \BibitemShut {NoStop}%
\bibitem [{\citenamefont {Halverson}\ \emph {et~al.}(2019)\citenamefont {Halverson}, \citenamefont {Long}, \citenamefont {Nelson},\ and\ \citenamefont {Salinas}}]{Halverson:2019cmy}%
  \BibitemOpen
  \bibfield  {author} {\bibinfo {author} {\bibfnamefont {J.}~\bibnamefont {Halverson}}, \bibinfo {author} {\bibfnamefont {C.}~\bibnamefont {Long}}, \bibinfo {author} {\bibfnamefont {B.}~\bibnamefont {Nelson}},\ and\ \bibinfo {author} {\bibfnamefont {G.}~\bibnamefont {Salinas}},\ }\href {https://doi.org/10.1103/PhysRevD.100.106010} {\bibfield  {journal} {\bibinfo  {journal} {Phys. Rev. D}\ }\textbf {\bibinfo {volume} {100}},\ \bibinfo {pages} {106010} (\bibinfo {year} {2019})},\ \Eprint {https://arxiv.org/abs/1909.05257} {arXiv:1909.05257 [hep-th]} \BibitemShut {NoStop}%
\bibitem [{\citenamefont {Witten}(1984)}]{Witten:1984dg}%
  \BibitemOpen
  \bibfield  {author} {\bibinfo {author} {\bibfnamefont {E.}~\bibnamefont {Witten}},\ }\href {https://doi.org/10.1016/0370-2693(84)90422-2} {\bibfield  {journal} {\bibinfo  {journal} {Phys. Lett. B}\ }\textbf {\bibinfo {volume} {149}},\ \bibinfo {pages} {351} (\bibinfo {year} {1984})}\BibitemShut {NoStop}%
\bibitem [{\citenamefont {Tegmark}\ \emph {et~al.}(2006{\natexlab{b}})\citenamefont {Tegmark} \emph {et~al.}}]{PhysRevD.73.023505}%
  \BibitemOpen
  \bibfield  {author} {\bibinfo {author} {\bibfnamefont {M.}~\bibnamefont {Tegmark}} \emph {et~al.},\ }\href {https://doi.org/10.1103/PhysRevD.73.023505} {\bibfield  {journal} {\bibinfo  {journal} {Phys. Rev. D}\ }\textbf {\bibinfo {volume} {73}},\ \bibinfo {pages} {023505} (\bibinfo {year} {2006}{\natexlab{b}})}\BibitemShut {NoStop}%
\bibitem [{\citenamefont {Sikivie}(1983)}]{Sikivie1983}%
  \BibitemOpen
  \bibfield  {author} {\bibinfo {author} {\bibfnamefont {P.}~\bibnamefont {Sikivie}},\ }\href {https://doi.org/10.1103/PhysRevLett.51.1415} {\bibfield  {journal} {\bibinfo  {journal} {Phys.~Rev.~Lett.}\ }\textbf {\bibinfo {volume} {51}},\ \bibinfo {pages} {1415} (\bibinfo {year} {1983})},\ \bibinfo {note} {[Erratum: Phys.~Rev.~Lett.~52, 695 (1984)]}\BibitemShut {NoStop}%
\bibitem [{\citenamefont {de~Salas}\ and\ \citenamefont {Widmark}(2021)}]{de_Salas_2021}%
  \BibitemOpen
  \bibfield  {author} {\bibinfo {author} {\bibfnamefont {P.~F.}\ \bibnamefont {de~Salas}}\ and\ \bibinfo {author} {\bibfnamefont {A.}~\bibnamefont {Widmark}},\ }\href {https://doi.org/10.1088/1361-6633/ac24e7} {\bibfield  {journal} {\bibinfo  {journal} {Reports on Progress in Physics}\ }\textbf {\bibinfo {volume} {84}},\ \bibinfo {pages} {104901} (\bibinfo {year} {2021})},\ \Eprint {https://arxiv.org/abs/2012.11477} {arXiv:2012.11477 [astro-ph.GA]} \BibitemShut {NoStop}%
\bibitem [{\citenamefont {Bartram}\ \emph {et~al.}(2021{\natexlab{a}})\citenamefont {Bartram} \emph {et~al.}}]{PhysRevLett.127.261803}%
  \BibitemOpen
  \bibfield  {author} {\bibinfo {author} {\bibfnamefont {C.}~\bibnamefont {Bartram}} \emph {et~al.} (\bibinfo {collaboration} {ADMX Collaboration}),\ }\href {https://doi.org/10.1103/PhysRevLett.127.261803} {\bibfield  {journal} {\bibinfo  {journal} {Phys. Rev. Lett.}\ }\textbf {\bibinfo {volume} {127}},\ \bibinfo {pages} {261803} (\bibinfo {year} {2021}{\natexlab{a}})},\ \Eprint {https://arxiv.org/abs/2110.06096} {arXiv:2110.06096 [hep-ex]} \BibitemShut {NoStop}%
\bibitem [{\citenamefont {Backes}\ \emph {et~al.}(2021)\citenamefont {Backes} \emph {et~al.}}]{Backes2021}%
  \BibitemOpen
  \bibfield  {author} {\bibinfo {author} {\bibfnamefont {K.~M.}\ \bibnamefont {Backes}} \emph {et~al.} (\bibinfo {collaboration} {HAYSTAC Collaboration}),\ }\href {https://doi.org/10.1038/s41586-021-03226-7} {\bibfield  {journal} {\bibinfo  {journal} {Nature}\ }\textbf {\bibinfo {volume} {590}},\ \bibinfo {pages} {238} (\bibinfo {year} {2021})},\ \Eprint {https://arxiv.org/abs/2008.01853} {arXiv:2008.01853 [quant-ph]} \BibitemShut {NoStop}%
\bibitem [{\citenamefont {Yi}\ \emph {et~al.}(2023)\citenamefont {Yi} \emph {et~al.}}]{Yi:2022fmn}%
  \BibitemOpen
  \bibfield  {author} {\bibinfo {author} {\bibfnamefont {A.~K.}\ \bibnamefont {Yi}} \emph {et~al.},\ }\href {https://doi.org/10.1103/PhysRevLett.130.071002} {\bibfield  {journal} {\bibinfo  {journal} {Phys. Rev. Lett.}\ }\textbf {\bibinfo {volume} {130}},\ \bibinfo {pages} {071002} (\bibinfo {year} {2023})},\ \Eprint {https://arxiv.org/abs/2210.10961} {arXiv:2210.10961 [hep-ex]} \BibitemShut {NoStop}%
\bibitem [{\citenamefont {Cabrera}\ and\ \citenamefont {Thomas}(2008)}]{Cabrera2008}%
  \BibitemOpen
  \bibfield  {author} {\bibinfo {author} {\bibfnamefont {B.}~\bibnamefont {Cabrera}}\ and\ \bibinfo {author} {\bibfnamefont {S.}~\bibnamefont {Thomas}},\ }\href {http://www.physics.rutgers.edu/~scthomas/talks/Axion-LC-Florida.pdf} {\bibinfo {title} {{Workshop Axions 2010, U. Florida}}} (\bibinfo {year} {2008})\BibitemShut {NoStop}%
\bibitem [{\citenamefont {Sikivie}\ \emph {et~al.}(2014)\citenamefont {Sikivie} \emph {et~al.}}]{sikivie2014proposal}%
  \BibitemOpen
  \bibfield  {author} {\bibinfo {author} {\bibfnamefont {P.}~\bibnamefont {Sikivie}} \emph {et~al.},\ }\href {https://doi.org/10.1103/PhysRevLett.112.131301} {\bibfield  {journal} {\bibinfo  {journal} {Physical Review Letters}\ }\textbf {\bibinfo {volume} {112}},\ \bibinfo {pages} {131301} (\bibinfo {year} {2014})},\ \Eprint {https://arxiv.org/abs/1310.8545} {arXiv:1310.8545 [hep-ph]} \BibitemShut {NoStop}%
\bibitem [{\citenamefont {Crisosto}\ \emph {et~al.}(2020)\citenamefont {Crisosto} \emph {et~al.}}]{Crisosto:2019fcj}%
  \BibitemOpen
  \bibfield  {author} {\bibinfo {author} {\bibfnamefont {N.}~\bibnamefont {Crisosto}} \emph {et~al.},\ }\href {https://doi.org/10.1103/PhysRevLett.124.241101} {\bibfield  {journal} {\bibinfo  {journal} {Phys. Rev. Lett.}\ }\textbf {\bibinfo {volume} {124}},\ \bibinfo {pages} {241101} (\bibinfo {year} {2020})},\ \Eprint {https://arxiv.org/abs/1911.05772} {arXiv:1911.05772 [astro-ph.CO]} \BibitemShut {NoStop}%
\bibitem [{\citenamefont {Ouellet}\ \emph {et~al.}(2019{\natexlab{a}})\citenamefont {Ouellet} \emph {et~al.}}]{PhysRevLett.122.121802}%
  \BibitemOpen
  \bibfield  {author} {\bibinfo {author} {\bibfnamefont {J.~L.}\ \bibnamefont {Ouellet}} \emph {et~al.},\ }\href {https://doi.org/10.1103/PhysRevLett.122.121802} {\bibfield  {journal} {\bibinfo  {journal} {Phys.~Rev.~Lett.}\ }\textbf {\bibinfo {volume} {122}},\ \bibinfo {pages} {121802} (\bibinfo {year} {2019}{\natexlab{a}})},\ \Eprint {https://arxiv.org/abs/1810.12257} {arXiv:1810.12257 [hep-ex]} \BibitemShut {NoStop}%
\bibitem [{\citenamefont {Ouellet}\ \emph {et~al.}(2019{\natexlab{b}})\citenamefont {Ouellet} \emph {et~al.}}]{PhysRevD.99.052012}%
  \BibitemOpen
  \bibfield  {author} {\bibinfo {author} {\bibfnamefont {J.~L.}\ \bibnamefont {Ouellet}} \emph {et~al.},\ }\href {https://doi.org/10.1103/PhysRevD.99.052012} {\bibfield  {journal} {\bibinfo  {journal} {Phys.~Rev.~D}\ }\textbf {\bibinfo {volume} {99}},\ \bibinfo {pages} {052012} (\bibinfo {year} {2019}{\natexlab{b}})},\ \Eprint {https://arxiv.org/abs/1901.10652} {arXiv:1901.10652 [hep-ex]} \BibitemShut {NoStop}%
\bibitem [{\citenamefont {Salemi}\ \emph {et~al.}(2021)\citenamefont {Salemi} \emph {et~al.}}]{PhysRevLett.127.081801}%
  \BibitemOpen
  \bibfield  {author} {\bibinfo {author} {\bibfnamefont {C.~P.}\ \bibnamefont {Salemi}} \emph {et~al.},\ }\href {https://doi.org/10.1103/PhysRevLett.127.081801} {\bibfield  {journal} {\bibinfo  {journal} {Phys.~Rev.~Lett.}\ }\textbf {\bibinfo {volume} {127}},\ \bibinfo {pages} {081801} (\bibinfo {year} {2021})},\ \Eprint {https://arxiv.org/abs/2102.06722} {arXiv:2102.06722 [hep-ex]} \BibitemShut {NoStop}%
\bibitem [{\citenamefont {Kahn}\ \emph {et~al.}(2016)\citenamefont {Kahn} \emph {et~al.}}]{PhysRevLett.117.141801}%
  \BibitemOpen
  \bibfield  {author} {\bibinfo {author} {\bibfnamefont {Y.}~\bibnamefont {Kahn}} \emph {et~al.},\ }\href {https://doi.org/10.1103/PhysRevLett.117.141801} {\bibfield  {journal} {\bibinfo  {journal} {Phys. Rev. Lett.}\ }\textbf {\bibinfo {volume} {117}},\ \bibinfo {pages} {141801} (\bibinfo {year} {2016})},\ \Eprint {https://arxiv.org/abs/1602.01086} {arXiv:1602.01086 [hep-ph]} \BibitemShut {NoStop}%
\bibitem [{\citenamefont {Gramolin}\ \emph {et~al.}(2021)\citenamefont {Gramolin} \emph {et~al.}}]{Gramolin2020a}%
  \BibitemOpen
  \bibfield  {author} {\bibinfo {author} {\bibfnamefont {A.~V.}\ \bibnamefont {Gramolin}} \emph {et~al.},\ }\href {https://doi.org/10.1038/s41567-020-1006-6} {\bibfield  {journal} {\bibinfo  {journal} {Nature Physics}\ }\textbf {\bibinfo {volume} {17}},\ \bibinfo {pages} {79} (\bibinfo {year} {2021})},\ \Eprint {https://arxiv.org/abs/2003.03348} {arXiv:2003.03348 [hep-ex]} \BibitemShut {NoStop}%
\bibitem [{\citenamefont {Rapidis}(2022)}]{Rapidis:2022proceedings}%
  \BibitemOpen
  \bibfield  {author} {\bibinfo {author} {\bibfnamefont {N.~M.}\ \bibnamefont {Rapidis}} (\bibinfo {collaboration} {DMRadio Collaboration}),\ }\href@noop {} {\  (\bibinfo {year} {2022})},\ \Eprint {https://arxiv.org/abs/2210.07215} {arXiv:2210.07215 [hep-ex]} \BibitemShut {NoStop}%
\bibitem [{\citenamefont {Phipps}\ \emph {et~al.}(2020)\citenamefont {Phipps} \emph {et~al.}}]{10.1007/978-3-030-43761-9_16}%
  \BibitemOpen
  \bibfield  {author} {\bibinfo {author} {\bibfnamefont {A.}~\bibnamefont {Phipps}} \emph {et~al.},\ }in\ \href {https://doi.org/10.1007/978-3-030-43761-9_16} {\emph {\bibinfo {booktitle} {Microwave Cavities and Detectors for Axion Research}}},\ \bibinfo {editor} {edited by\ \bibinfo {editor} {\bibfnamefont {G.}~\bibnamefont {Carosi}}\ and\ \bibinfo {editor} {\bibfnamefont {G.}~\bibnamefont {Rybka}}}\ (\bibinfo  {publisher} {Springer International Publishing},\ \bibinfo {address} {Cham},\ \bibinfo {year} {2020})\ pp.\ \bibinfo {pages} {139--145}\BibitemShut {NoStop}%
\bibitem [{\citenamefont {Benabou}\ \emph {et~al.}(2022)\citenamefont {Benabou} \emph {et~al.}}]{Benabou:2022arx}%
  \BibitemOpen
  \bibfield  {author} {\bibinfo {author} {\bibfnamefont {J.}~\bibnamefont {Benabou}} \emph {et~al.},\ }\href@noop {} {\  (\bibinfo {year} {2022})},\ \Eprint {https://arxiv.org/abs/2211.00008} {arXiv:2211.00008 [hep-ph]} \BibitemShut {NoStop}%
\bibitem [{\citenamefont {Budker}\ \emph {et~al.}(2014)\citenamefont {Budker} \emph {et~al.}}]{Budker:2013hfa}%
  \BibitemOpen
  \bibfield  {author} {\bibinfo {author} {\bibfnamefont {D.}~\bibnamefont {Budker}} \emph {et~al.},\ }\href {https://doi.org/10.1103/PhysRevX.4.021030} {\bibfield  {journal} {\bibinfo  {journal} {Phys. Rev. X}\ }\textbf {\bibinfo {volume} {4}},\ \bibinfo {pages} {021030} (\bibinfo {year} {2014})},\ \Eprint {https://arxiv.org/abs/1306.6089} {arXiv:1306.6089 [hep-ph]} \BibitemShut {NoStop}%
\bibitem [{\citenamefont {Bogorad}\ \emph {et~al.}(2019)\citenamefont {Bogorad} \emph {et~al.}}]{PhysRevLett.123.021801}%
  \BibitemOpen
  \bibfield  {author} {\bibinfo {author} {\bibfnamefont {Z.}~\bibnamefont {Bogorad}} \emph {et~al.},\ }\href {https://doi.org/10.1103/PhysRevLett.123.021801} {\bibfield  {journal} {\bibinfo  {journal} {Phys. Rev. Lett.}\ }\textbf {\bibinfo {volume} {123}},\ \bibinfo {pages} {021801} (\bibinfo {year} {2019})},\ \Eprint {https://arxiv.org/abs/1902.01418} {arXiv:1902.01418 [hep-ph]} \BibitemShut {NoStop}%
\bibitem [{\citenamefont {Brouwer}\ \emph {et~al.}(2022{\natexlab{a}})\citenamefont {Brouwer} \emph {et~al.}}]{Brouwer:2022DMRm}%
  \BibitemOpen
  \bibfield  {author} {\bibinfo {author} {\bibfnamefont {L.}~\bibnamefont {Brouwer}} \emph {et~al.} (\bibinfo {collaboration} {DMRadio Collaboration}),\ }\href {https://doi.org/10.1103/PhysRevD.106.103008} {\bibfield  {journal} {\bibinfo  {journal} {Phys. Rev. D}\ }\textbf {\bibinfo {volume} {106}},\ \bibinfo {pages} {103008} (\bibinfo {year} {2022}{\natexlab{a}})},\ \Eprint {https://arxiv.org/abs/2204.13781} {arXiv:2204.13781 [hep-ex]} \BibitemShut {NoStop}%
\bibitem [{\citenamefont {Chaudhuri}\ \emph {et~al.}(2018)\citenamefont {Chaudhuri} \emph {et~al.}}]{Chaudhuri:2018rqn}%
  \BibitemOpen
  \bibfield  {author} {\bibinfo {author} {\bibfnamefont {S.}~\bibnamefont {Chaudhuri}} \emph {et~al.},\ }\href@noop {} {\  (\bibinfo {year} {2018})},\ \Eprint {https://arxiv.org/abs/1803.01627} {arXiv:1803.01627 [hep-ph]} \BibitemShut {NoStop}%
\bibitem [{\citenamefont {Dine}\ \emph {et~al.}(1981)\citenamefont {Dine} \emph {et~al.}}]{Dine:1981rt}%
  \BibitemOpen
  \bibfield  {author} {\bibinfo {author} {\bibfnamefont {M.}~\bibnamefont {Dine}} \emph {et~al.},\ }\href {https://doi.org/10.1016/0370-2693(81)90590-6} {\bibfield  {journal} {\bibinfo  {journal} {Phys.~Lett.~B}\ }\textbf {\bibinfo {volume} {104}},\ \bibinfo {pages} {199} (\bibinfo {year} {1981})}\BibitemShut {NoStop}%
\bibitem [{\citenamefont {Zhitnitsky}(1980)}]{Zhitnitsky:1980tq}%
  \BibitemOpen
  \bibfield  {author} {\bibinfo {author} {\bibfnamefont {A.}~\bibnamefont {Zhitnitsky}},\ }\href {https://www.osti.gov/biblio/7063072} {\bibfield  {journal} {\bibinfo  {journal} {Sov.~J.~Nucl.~Phys.}\ }\textbf {\bibinfo {volume} {31}},\ \bibinfo {pages} {260} (\bibinfo {year} {1980})}\BibitemShut {NoStop}%
\bibitem [{\citenamefont {Kim}(1979)}]{PhysRevLett.43.103}%
  \BibitemOpen
  \bibfield  {author} {\bibinfo {author} {\bibfnamefont {J.~E.}\ \bibnamefont {Kim}},\ }\href {https://doi.org/10.1103/PhysRevLett.43.103} {\bibfield  {journal} {\bibinfo  {journal} {Phys. Rev. Lett.}\ }\textbf {\bibinfo {volume} {43}},\ \bibinfo {pages} {103} (\bibinfo {year} {1979})}\BibitemShut {NoStop}%
\bibitem [{\citenamefont {Shifman}\ \emph {et~al.}(1980)\citenamefont {Shifman} \emph {et~al.}}]{SHIFMAN1980493}%
  \BibitemOpen
  \bibfield  {author} {\bibinfo {author} {\bibfnamefont {M.}~\bibnamefont {Shifman}} \emph {et~al.},\ }\href {https://doi.org/https://doi.org/10.1016/0550-3213(80)90209-6} {\bibfield  {journal} {\bibinfo  {journal} {Nucl.~Phys.~B}\ }\textbf {\bibinfo {volume} {166}},\ \bibinfo {pages} {493 } (\bibinfo {year} {1980})}\BibitemShut {NoStop}%
\bibitem [{\citenamefont {Pozar}(2011)}]{pozar2011microwave}%
  \BibitemOpen
  \bibfield  {author} {\bibinfo {author} {\bibfnamefont {D.~M.}\ \bibnamefont {Pozar}},\ }\href@noop {} {\emph {\bibinfo {title} {Microwave Engineering}}}\ (\bibinfo  {publisher} {John Wiley \& Sons},\ \bibinfo {year} {2011})\BibitemShut {NoStop}%
\bibitem [{\citenamefont {{COMSOL AB}}()}]{comsol}%
  \BibitemOpen
  \bibfield  {author} {\bibinfo {author} {\bibnamefont {{COMSOL AB}}},\ }\href {https://www.comsol.com} {\bibinfo {title} {Comsol multiphysics}}\BibitemShut {NoStop}%
\bibitem [{\citenamefont {Braine}\ \emph {et~al.}(2020)\citenamefont {Braine} \emph {et~al.}}]{ADMX:2019uok}%
  \BibitemOpen
  \bibfield  {author} {\bibinfo {author} {\bibfnamefont {T.}~\bibnamefont {Braine}} \emph {et~al.} (\bibinfo {collaboration} {ADMX Collaboration}),\ }\href {https://doi.org/10.1103/PhysRevLett.124.101303} {\bibfield  {journal} {\bibinfo  {journal} {Phys. Rev. Lett.}\ }\textbf {\bibinfo {volume} {124}},\ \bibinfo {pages} {101303} (\bibinfo {year} {2020})},\ \Eprint {https://arxiv.org/abs/1910.08638} {arXiv:1910.08638 [hep-ex]} \BibitemShut {NoStop}%
\bibitem [{\citenamefont {Brubaker}\ \emph {et~al.}(2017)\citenamefont {Brubaker} \emph {et~al.}}]{Brubaker2017}%
  \BibitemOpen
  \bibfield  {author} {\bibinfo {author} {\bibfnamefont {B.~M.}\ \bibnamefont {Brubaker}} \emph {et~al.},\ }\href {https://doi.org/https://doi.org/10.1103/PhysRevLett.118.061302} {\bibfield  {journal} {\bibinfo  {journal} {Phys. Rev. Lett.}\ }\textbf {\bibinfo {volume} {118}},\ \bibinfo {pages} {061302} (\bibinfo {year} {2017})},\ \Eprint {https://arxiv.org/abs/1610.02580} {arXiv:1610.02580 [astro-ph.CO]} \BibitemShut {NoStop}%
\bibitem [{\citenamefont {Rapidis}\ \emph {et~al.}(2019)\citenamefont {Rapidis} \emph {et~al.}}]{Rapidis:2018dci}%
  \BibitemOpen
  \bibfield  {author} {\bibinfo {author} {\bibfnamefont {N.~M.}\ \bibnamefont {Rapidis}} \emph {et~al.},\ }\href {https://doi.org/10.1063/1.5055246} {\bibfield  {journal} {\bibinfo  {journal} {Rev. Sci. Instrum.}\ }\textbf {\bibinfo {volume} {90}},\ \bibinfo {pages} {024706} (\bibinfo {year} {2019})},\ \Eprint {https://arxiv.org/abs/1809.02246} {arXiv:1809.02246 [physics.ins-det]} \BibitemShut {NoStop}%
\bibitem [{\citenamefont {Herzog-Arbeitman}\ \emph {et~al.}(2018)\citenamefont {Herzog-Arbeitman} \emph {et~al.}}]{Herzog-Arbeitman:2017fte}%
  \BibitemOpen
  \bibfield  {author} {\bibinfo {author} {\bibfnamefont {J.}~\bibnamefont {Herzog-Arbeitman}} \emph {et~al.},\ }\href {https://doi.org/10.1103/PhysRevLett.120.041102} {\bibfield  {journal} {\bibinfo  {journal} {Phys. Rev. Lett.}\ }\textbf {\bibinfo {volume} {120}},\ \bibinfo {pages} {041102} (\bibinfo {year} {2018})},\ \Eprint {https://arxiv.org/abs/1704.04499} {arXiv:1704.04499 [astro-ph.GA]} \BibitemShut {NoStop}%
\bibitem [{\citenamefont {Brouwer}\ \emph {et~al.}(2022{\natexlab{b}})\citenamefont {Brouwer} \emph {et~al.}}]{DMRadio:2022jfv}%
  \BibitemOpen
  \bibfield  {author} {\bibinfo {author} {\bibfnamefont {L.}~\bibnamefont {Brouwer}} \emph {et~al.} (\bibinfo {collaboration} {DMRadio Collaboration}),\ }\href {https://doi.org/10.1103/PhysRevD.106.112003} {\bibfield  {journal} {\bibinfo  {journal} {Phys. Rev. D}\ }\textbf {\bibinfo {volume} {106}},\ \bibinfo {pages} {112003} (\bibinfo {year} {2022}{\natexlab{b}})},\ \Eprint {https://arxiv.org/abs/2203.11246} {arXiv:2203.11246 [hep-ex]} \BibitemShut {NoStop}%
\bibitem [{\citenamefont {Ankel}\ \emph {et~al.}(2025)\citenamefont {Ankel} \emph {et~al.}}]{DMRmSquids}%
  \BibitemOpen
  \bibfield  {author} {\bibinfo {author} {\bibfnamefont {V.}~\bibnamefont {Ankel}} \emph {et~al.},\ }\href@noop {} {\  (\bibinfo {year} {2025})},\ \Eprint {https://arxiv.org/abs/2504.20398} {arXiv:2504.20398 [quant-ph]} \BibitemShut {NoStop}%
\bibitem [{\citenamefont {AlShirawi}\ \emph {et~al.}(2025)\citenamefont {AlShirawi} \emph {et~al.}}]{alshirawi_2025_15149989}%
  \BibitemOpen
  \bibfield  {author} {\bibinfo {author} {\bibfnamefont {A.}~\bibnamefont {AlShirawi}} \emph {et~al.},\ }\href {https://doi.org/10.5281/zenodo.15149989} {10.5281/zenodo.15149989} (\bibinfo {year} {2025})\BibitemShut {NoStop}%
\bibitem [{\citenamefont {Bartram}\ \emph {et~al.}(2021{\natexlab{b}})\citenamefont {Bartram} \emph {et~al.}}]{ADMX:2020hay}%
  \BibitemOpen
  \bibfield  {author} {\bibinfo {author} {\bibfnamefont {C.}~\bibnamefont {Bartram}} \emph {et~al.} (\bibinfo {collaboration} {ADMX Collaboration}),\ }\href {https://doi.org/10.1103/PhysRevD.103.032002} {\bibfield  {journal} {\bibinfo  {journal} {Phys. Rev. D}\ }\textbf {\bibinfo {volume} {103}},\ \bibinfo {pages} {032002} (\bibinfo {year} {2021}{\natexlab{b}})},\ \Eprint {https://arxiv.org/abs/2010.06183} {arXiv:2010.06183 [astro-ph.CO]} \BibitemShut {NoStop}%
\bibitem [{\citenamefont {{ADMX Collaboration}}()}]{ADMXprivatecom}%
  \BibitemOpen
  \bibfield  {author} {\bibinfo {author} {\bibnamefont {{ADMX Collaboration}}},\ }\href@noop {} {}\bibinfo {howpublished} {Private Communication}\BibitemShut {NoStop}%
\bibitem [{\citenamefont {Palken}\ \emph {et~al.}(2020)\citenamefont {Palken} \emph {et~al.}}]{Palken:2020wgs}%
  \BibitemOpen
  \bibfield  {author} {\bibinfo {author} {\bibfnamefont {D.~A.}\ \bibnamefont {Palken}} \emph {et~al.},\ }\href {https://doi.org/10.1103/PhysRevD.101.123011} {\bibfield  {journal} {\bibinfo  {journal} {Phys. Rev. D}\ }\textbf {\bibinfo {volume} {101}},\ \bibinfo {pages} {123011} (\bibinfo {year} {2020})},\ \Eprint {https://arxiv.org/abs/2003.08510} {arXiv:2003.08510 [astro-ph.IM]} \BibitemShut {NoStop}%
\bibitem [{\citenamefont {Jackson}(1999)}]{jackson_classical_1999}%
  \BibitemOpen
  \bibfield  {author} {\bibinfo {author} {\bibfnamefont {J.~D.}\ \bibnamefont {Jackson}},\ }\href {http://cdsweb.cern.ch/record/490457} {\emph {\bibinfo {title} {Classical electrodynamics}}},\ \bibinfo {edition} {3rd}\ ed.\ (\bibinfo  {publisher} {Wiley},\ \bibinfo {address} {New York, {NY}},\ \bibinfo {year} {1999})\BibitemShut {NoStop}%
\bibitem [{\citenamefont {Pippard}(1954)}]{PIPPARD19541}%
  \BibitemOpen
  \bibfield  {author} {\bibinfo {author} {\bibfnamefont {A.}~\bibnamefont {Pippard}},\ }\href {https://doi.org/https://doi.org/10.1016/S0065-2539(08)60130-4} {\ \bibinfo {series} {Advances in Electronics and Electron Physics},\ \textbf {\bibinfo {volume} {6}},\ \bibinfo {pages} {1} (\bibinfo {year} {1954})}\BibitemShut {NoStop}%
\bibitem [{\citenamefont {Cahill}\ \emph {et~al.}(2016)\citenamefont {Cahill} \emph {et~al.}}]{Cahill:2016uui}%
  \BibitemOpen
  \bibfield  {author} {\bibinfo {author} {\bibfnamefont {A.}~\bibnamefont {Cahill}} \emph {et~al.},\ }in\ \href {https://doi.org/10.18429/JACoW-IPAC2016-MOPMW038} {\emph {\bibinfo {booktitle} {{7th International Particle Accelerator Conference}}}}\ (\bibinfo {year} {2016})\BibitemShut {NoStop}%
\bibitem [{\citenamefont {Peng}\ \emph {et~al.}(2000)\citenamefont {Peng} \emph {et~al.}}]{peng2000cryogenic}%
  \BibitemOpen
  \bibfield  {author} {\bibinfo {author} {\bibfnamefont {H.}~\bibnamefont {Peng}} \emph {et~al.},\ }\href {https://doi.org/https://doi.org/10.1016/S0168-9002(99)00971-7} {\bibfield  {journal} {\bibinfo  {journal} {Nuclear Instruments and Methods in Physics Research Section A: Accelerators, Spectrometers, Detectors and Associated Equipment}\ }\textbf {\bibinfo {volume} {444}},\ \bibinfo {pages} {569} (\bibinfo {year} {2000})}\BibitemShut {NoStop}%
\bibitem [{\citenamefont {Finger}\ and\ \citenamefont {Kerr}(2008)}]{finger2008microwave}%
  \BibitemOpen
  \bibfield  {author} {\bibinfo {author} {\bibfnamefont {R.}~\bibnamefont {Finger}}\ and\ \bibinfo {author} {\bibfnamefont {A.}~\bibnamefont {Kerr}},\ }\href {https://doi.org/10.1007/s10762-008-9394-1} {\bibfield  {journal} {\bibinfo  {journal} {International Journal of Infrared and Millimeter Waves}\ }\textbf {\bibinfo {volume} {29}},\ \bibinfo {pages} {924} (\bibinfo {year} {2008})},\ \Eprint {https://arxiv.org/abs/1509.05273} {arXiv:1509.05273 [astro-ph]} \BibitemShut {NoStop}%
\bibitem [{\citenamefont {Simanovskaia}\ \emph {et~al.}(2021)\citenamefont {Simanovskaia} \emph {et~al.}}]{simanovskaia2021}%
  \BibitemOpen
  \bibfield  {author} {\bibinfo {author} {\bibfnamefont {M.}~\bibnamefont {Simanovskaia}} \emph {et~al.},\ }\href {https://doi.org/10.1063/5.0016125} {\bibfield  {journal} {\bibinfo  {journal} {Review of Scientific Instruments}\ }\textbf {\bibinfo {volume} {92}},\ \bibinfo {pages} {033305} (\bibinfo {year} {2021})},\ \Eprint {https://arxiv.org/abs/2006.01248} {arXiv:2006.01248 [astro-ph]} \BibitemShut {NoStop}%
\bibitem [{\citenamefont {Yoon}\ \emph {et~al.}(2022)\citenamefont {Yoon} \emph {et~al.}}]{PhysRevD.106.092007}%
  \BibitemOpen
  \bibfield  {author} {\bibinfo {author} {\bibfnamefont {H.}~\bibnamefont {Yoon}} \emph {et~al.},\ }\href {https://doi.org/10.1103/PhysRevD.106.092007} {\bibfield  {journal} {\bibinfo  {journal} {Phys. Rev. D}\ }\textbf {\bibinfo {volume} {106}},\ \bibinfo {pages} {092007} (\bibinfo {year} {2022})},\ \Eprint {https://arxiv.org/abs/2206.12271} {arXiv:2206.12271 [hep-ex]} \BibitemShut {NoStop}%
\bibitem [{\citenamefont {Benz}(1969)}]{benz1969magnetoresistance}%
  \BibitemOpen
  \bibfield  {author} {\bibinfo {author} {\bibfnamefont {M.}~\bibnamefont {Benz}},\ }\href {https://aip.scitation.org/doi/10.1063/1.1657896} {\bibfield  {journal} {\bibinfo  {journal} {Journal of Applied Physics}\ }\textbf {\bibinfo {volume} {40}},\ \bibinfo {pages} {2003} (\bibinfo {year} {1969})}\BibitemShut {NoStop}%
\bibitem [{\citenamefont {Fickett}(1972)}]{fickett1972magnetoresistivity}%
  \BibitemOpen
  \bibfield  {author} {\bibinfo {author} {\bibfnamefont {F.~R.}\ \bibnamefont {Fickett}},\ }\href {https://lss.fnal.gov/conf/C720919/p539.pdf} {\bibfield  {journal} {\bibinfo  {journal} {eConf}\ }\textbf {\bibinfo {volume} {C720919}},\ \bibinfo {pages} {539} (\bibinfo {year} {1972})}\BibitemShut {NoStop}%
\bibitem [{\citenamefont {Kohler}(1938)}]{kohler1938magnetischen}%
  \BibitemOpen
  \bibfield  {author} {\bibinfo {author} {\bibfnamefont {M.}~\bibnamefont {Kohler}},\ }\href {https://doi.org/10.1002/andp.19384240124} {\bibfield  {journal} {\bibinfo  {journal} {Annalen der Physik}\ }\textbf {\bibinfo {volume} {424}},\ \bibinfo {pages} {211} (\bibinfo {year} {1938})}\BibitemShut {NoStop}%
\bibitem [{\citenamefont {Rogers}\ \emph {et~al.}(1988)\citenamefont {Rogers} \emph {et~al.}}]{rogers1988anomalous}%
  \BibitemOpen
  \bibfield  {author} {\bibinfo {author} {\bibfnamefont {J.}~\bibnamefont {Rogers}} \emph {et~al.},\ }\href {https://doi.org/10.1063/1.99509} {\bibfield  {journal} {\bibinfo  {journal} {Applied Physics Letters}\ }\textbf {\bibinfo {volume} {52}},\ \bibinfo {pages} {2266} (\bibinfo {year} {1988})}\BibitemShut {NoStop}%
\bibitem [{\citenamefont {Ahn}\ \emph {et~al.}(2017)\citenamefont {Ahn} \emph {et~al.}}]{ahn2017magnetoresistance}%
  \BibitemOpen
  \bibfield  {author} {\bibinfo {author} {\bibfnamefont {S.}~\bibnamefont {Ahn}} \emph {et~al.},\ }\href {https://dx.doi.org/10.1088/1748-0221/12/10/P10023} {\bibfield  {journal} {\bibinfo  {journal} {Journal of Instrumentation}\ }\textbf {\bibinfo {volume} {12}}\bibfield  {number} {\bibinfo  {number} { (10)},\ \bibinfo {pages} {P10023}},\ }\Eprint {https://arxiv.org/abs/1705.04754} {arXiv:1705.04754 [physics.ins-det]} \BibitemShut {NoStop}%
\end{thebibliography}%


\end{document}